\documentclass[a4paper,fleqn,usenatbib]{mnras}

\usepackage[T1]{fontenc}
\usepackage{ae,aecompl}

\pdfoutput=1
\usepackage{graphicx}	
\usepackage{amsmath}	
\usepackage{amssymb}	
\usepackage{booktabs} 
\usepackage{multirow} 
\usepackage{mathrsfs}

\usepackage{url}

\usepackage{siunitx} 
\DeclareSIUnit\year{yr}
\DeclareSIUnit\parsec{pc}
\DeclareSIUnit\erg{erg}
\DeclareSIUnit\janskey{Jy}
\DeclareSIUnit\beam{beam}
\DeclareSIUnit\msun{\ensuremath{\mathrm{M_{\sun}}}}
\DeclareSIUnit\lsun{\ensuremath{\mathrm{L_{\sun}}}}
\DeclareSIUnit\zsun{\ensuremath{\mathrm{Z_{\sun}}}}
\DeclareSIUnit\rsun{\ensuremath{\mathrm{R_{\sun}}}}
\DeclareSIUnit\photons{phot}

\usepackage[acronym, nomain]{glossaries}
\makeglossaries

\usepackage[sharp]{easylist} 
\usepackage[caption=false]{subfig}

\usepackage[noabbrev]{cleveref} 
\creflabelformat{equation}{#2#1#3}

\newcommand{\customdot}{\boldsymbol{\cdot}}

\newcommand{\customvec}{\mathbfit}
\newcommand{\custommat}{\mathbfss}
\newcommand{\pvec}[1]{\customvec{#1}\mkern2mu\vphantom{#1}} 
\newcommand{\mathpunc}[1] {\, #1} 

\newcommand{\upperroman}[1]{\uppercase\expandafter{\romannumeral#1}}

\newcommand\TopStrut{\rule{0pt}{2.6ex}}       
\newcommand\BottomStrut{\rule[-1.2ex]{0pt}{0pt}} 

\DeclareRobustCommand{\VAN}[3]{#2}

\newcommand\hii{H\textsc{ii}}
\newcommand\titlehii{H{\mdseries\textsc{ii}}}

\title[Cometary HII regions]{Hydrodynamical models of cometary \titlehii{} regions}

\author[H. G. Steggles et al.]{
H. G. Steggles,$^{1}$\thanks{E-mail: hgsteggles@gmail.com}
M. G. Hoare$^{1}$
and J. M. Pittard$^{1}$
\\
$^{1}$School of Physics and Astronomy, University of Leeds, Woodhouse Lane, Leeds LS2 9JT, UK\\
}

\date{Accepted 2016 December 23. Received 2016 December 22; in original form 2016 August 23}

\pubyear{2016}
\volume{466}

\begin{document}

\setacronymstyle{short-long}

\newacronym{rms}{RMS}{Red MSX Source}
\newacronym{torch}{TORCH}{The Operator-split Radiation, Cooling, Hydro-code}
\newacronym{ism}{ISM}{interstellar medium}
\glsunset{ism}
\newglossaryentry{uchii}
{
	name=UC\hii{},
	description={ultra-compact \hii{}},
	first={UC\hii{} (ultra-compact \hii{})}
}
\newacronym{amr}{AMR}{adaptive mesh refinement}
\newacronym{hllc}{HLLC}{Harten-Lax-van Leer-Contact}
\newacronym{hll}{HLL}{Harten-Lax-van Leer}
\newacronym{vla}{VLA}{Very Large Array}
\glsunset{vla}
\newacronym{cornish}{CORNISH}{Co-Ordinated Radio `N' Infrared Survey for High-mass star formation}
\glsunset{cornish}
\newacronym{tvd}{TVD}{total variation diminishing}
\newacronym{euv}{EUV}{extreme ultra-violet}
\newacronym{fuv}{FUV}{far ultra-violet}
\newacronym{fvm}{FVM}{finite-volume method}
\newacronym{lte}{LTE}{local thermodynamic equilibrium}
\newacronym{ir}{IR}{infra-red}
\glsunset{ir}

\label{firstpage}
\pagerange{4573--4591}
\maketitle

\begin{abstract}
We have modelled the evolution of cometary \hii{} regions produced by zero-age main-sequence stars of O and B spectral types, which are driving strong winds and are born off-centre from spherically symmetric cores with power-law ($\alpha = 2$) density slopes.
A model parameter grid was produced that spans stellar mass, age and core density.
Exploring this parameter space we investigated limb-brightening, a feature commonly seen in cometary \hii{} regions.
We found that stars with mass $M_\star \geq \SI{12}{\msun}$ produce this feature.
Our models have a cavity bounded by a contact discontinuity separating hot shocked wind and ionised ambient gas that is similar in size to the surrounding \hii{} region.
Due to early pressure confinement we did not see shocks outside of the contact discontinuity for stars with $M_\star \leq \SI{40}{\msun}$, but the cavities were found to continue to grow.
The cavity size in each model plateaus as the \hii{} region stagnates.
The spectral energy distributions of our models are similar to those from identical stars evolving in uniform density fields.
The turn-over frequency is slightly lower in our power-law models due to a higher proportion of low density gas covered by the \hii{} regions.
\end{abstract}

\glsreset{cornish}




\section{Introduction}

Massive stars ($M_\star > \SI{8}{\msun}$) are an important component of the universe. 
They are the main sources of stellar feedback that influences the structure, evolution and chemical composition of a galaxy.
High-mass stars are born in dense parts of the \gls*{ism}, erupting with outflows and radiatively driven winds that contribute to the destruction of the natal cloud.
Extreme ultra-violet radiation is emitted from high-mass stars, which ionises and heats the surrounding gas.
This produces an over-pressure that drives a shock into the ambient medium, sweeping gas into a shell that ultimately disperses the parent molecular cloud and may also trigger further star formation \citep{2007prpl.conf..181H}.
The optical and near-infrared emission from such stars is not able to penetrate the natal clouds due to high column densities of dust.
Radio free-free emission can, however, be seen right across the Galaxy \citep{2002ARA&A..40...27C}.
At wavelengths in the far-infrared, emission coming from warm dust re-emitting the stellar radiation can also be observed from ionised nebulae \citep{2015A&A...579A..71C}.

How high-mass stars are formed is still poorly understood.
They have short main-sequence lifetimes ($\lesssim \SI[retain-unity-mantissa=false]{1e8}{\year}$) which, along with their relatively low formation rates, indicates that only a small proportion ($\simeq \num[retain-unity-mantissa=false]{1e-6}$) of stars in the galaxy have high masses.
The scarcity of observational data is a significant problem for the testing of theoretical models; some important phases of evolution pass very quickly.
The youngest massive stars, however, are associated with \gls*{uchii} regions, a phase that lasts longer than expected \citep{1989ApJS...69..831W}.
\gls*{uchii} regions were defined by \citet{1989ApJS...69..831W} as photo-ionised nebulae with diameters \SI{\lesssim 0.1}{\parsec}, electron number densities \SI[retain-unity-mantissa=false]{\gtrsim 1e4}{\per\centi\metre\cubed} and emission measures \SI[retain-unity-mantissa=false]{\gtrsim 1e7}{\parsec\per\centi\metre\tothe{6}} that have not yet expanded out of their natal molecular cloud. 
The observed ionisation from such stars indicates that the ionising photon emission rates lie approximately between \SIlist[retain-unity-mantissa=false]{1e44;1e49}{\per\second} corresponding to ZAMS stars with spectral types B2--O5.
\citet{1989ApJ...340..265W} estimated the lifetime of \gls*{uchii} regions (\SI{\sim 4e5}{\year}) using the fraction of main-sequence stars observed in the \gls*{uchii} phase and an adopted value for the main-sequence lifetime (\SI{\sim 2.4e6}{\year}).
This lifetime is an order of magnitude longer than that predicted by simple Str\"{o}mgren sphere expansion (\SI{\sim 4e4}{\year}).

\gls*{uchii} regions are interesting to study as they may provide a lot of information about the early life of massive stars including their properties and, due to the limited number of observed morphological classes, their environments.
The \citet{1989ApJS...69..831W} survey included classifications of \gls*{uchii} regions into morphological types: cometary (\SI{\sim 20}{\percent}), core-halo (\SI{\sim 16}{\percent}), shell (\SI{\sim 4}{\percent}), irregular or multiple peaked structures (\SI{\sim 17}{\percent}) and spherical and unresolved (\SI{\sim 43}{\percent}).
\citet{1994ApJS...91..659K} made radio-continuum observations of 75 \gls*{uchii} regions for which the proportions of morphological types agree remarkably well with those in \citet{1989ApJS...69..831W}.
Observed at higher spatial resolution, some apparently spherical morphologies uncover more ordered morphologies \citep{1984A&A...136...53F}.
The radio continuum survey of \gls*{uchii} regions by \citet{1998MNRAS.301..640W} found that most sources were either cometary or irregular (ignoring the unresolved sources).
\citet{2005ApJ...624L.101D} revised the morphological scheme by adding the bipolar classification and removing the core-halo morphology.
The latter was due to the fact that essentially all \gls*{uchii} regions were observed to be associated with large-scale diffuse emission that could be separated from the compact cores.
The proportions of the morphologies in this survey compare well with the \citet{1989ApJS...69..831W} and \citet{1994ApJS...91..659K} surveys except that \SI{28}{\percent} of sources were classed as shell-like.
\citet{2007prpl.conf..181H} reviewed the different morphological surveys and noted that a lot of morphologies classified as shell-like could also be classified as cometary.

As \gls*{uchii} regions have had little time to significantly alter their natal environments, the morphologies that arise are thought to reflect the ambient density field.
Cometary types in particular are interesting to study and to test numerical models against because they have a highly regular shape and can overcome the lifetime problem.
In the past there were three major models that aimed to explain how \gls*{uchii} regions form cometary shapes: the champagne flow model \citep{1978A&A....70..769I}, the bow shock model \citep{1985ApJ...288L..17R, 1991ApJ...369..395M, 1992ApJ...394..534V} and clumpy/mass-loading models \citep{1995MNRAS.277..700D, 1996MNRAS.280..661R, 1996MNRAS.280..667W}.
A more recent idea says that the situation is probably best described by a combination of these i.e. hybrid models \citep{2006ApJS..165..283A}.

Cometary \hii{} regions are mostly observed at the edge of molecular clouds. 
Along with the shape, this is the reason they were first named ``blisters'' \citep{1978A&A....70..769I}.
This led to the champagne flow model, a modification of the simple Str\"{o}mgren solution in which the ionising star lies in an inhomogeneous (not a uniform) density field.
Early modelling by \citet{1979A&A....71...59T}, \citet{1979MNRAS.186...59W}, \citet{1979ApJ...233...85B}, \citet{1981A&A....98...85B} and \citet{1979A&A....80..110T} included discontinuities in density separating molecular clouds with the ISM.
\citet{1979ApJ...233...85B} found that once the ionised gas reaches the edge of the dense cloud, it accelerates up to \SI{\sim 30}{\kilo\metre\per\second}.
More realistically, \citet{1979A&A....78..352I}, \citet{1979ApJ...234..615I}, \citet{1980ApJ...236..808I} and \citet{1983A&A...127..313Y} modelled stars in density gradients, clearly reproducing cometary morphologies and predicting double spectral line profiles corresponding to a velocity splitting of up to \SI{50}{\kilo\meter\per\second} along the symmetry axis.
\citet{2005ApJ...627..813H} carried out simulations of the photo-evaporation of a cloud with large-scale density gradients.
They showed that an ionised flow is set up that has a transient phase with duration \SI[retain-unity-mantissa=false]{\sim 1e4}{\year}, which then becomes approximately stationary with respect to the ionisation front (the quasi-stationary phase) for a large part of its evolution.

Real \hii{} regions cannot be explained entirely in terms of a density gradient. 
Not only does the champagne model fail to exhibit limb-brightened morphologies, but also the lifetime problem is not solved due to the unconstrained expansion of the ionisation front. 
It also does not take into account that OB stars drive stellar winds that can significantly affect the dynamics of nearby gas.

If an ionising star moves supersonically with respect to a uniform density field, the strong stellar wind driven by that star balances the ram pressure produced by the ambient flow, forming a bow shock ahead of the star.
This model was introduced by \citet{1990ApJ...353..570V} as an alternative explanation of cometary \hii{} regions and later analysed numerically by \citet{1992ApJ...394..534V}.
They found that behind the bow shock neutral material is swept up into a very dense, thin shell that traps the expanding ionisation front.
This was a good result for the bow shock model as it was known that a confining mechanism was necessary to solve the lifetime problem. 
Limb-brightening was also found to occur in this model and can be explained as ionised gas that has been compressed by the stellar wind into a thin shell, flowing around the stellar wind cavity.
The bow shock model was attractive because the size and shape compared well with observed cometary \gls*{uchii} regions and it explained velocity gradients commonly seen at the head of cometaries.

\citet{2015ApJ...812...87Z} simulated the evolution of cometary \hii{} regions in bow shock and champagne flow models and analysed the resulting [Ne\textsc{ii}] \SI{12.81}{\micro\meter} and H\textsubscript{2}S(2) line profiles.
They discovered that the two models can be distinguished using these lines.

Both the bow shock model and the champagne flow model have been found to be inadequate in explaining the recombination line velocity structures of cometary \hii{} regions.
\citet{1996ApJ...464..272L} investigated the validity of both models in reproducing the velocity structure of the cometary \gls*{uchii} region G29.96--0.02 and found that individually they poorly describe the object.
\citet{1994ApJ...432..648G} suggested that by including stellar winds and non-uniform ambient density fields more realistic models could be made.

\citet{2007ApJ...660.1296F} studied the evolution of \hii{} regions moving from the centre to the edge of structured molecular clouds with Gaussian cores surrounded by power-law ($\rho \propto r^{-2}$ and $\rho \propto r^{-3}$) halos.
They found that \gls*{uchii} regions remain pressure-confined while they are inside the core but evolve into extended \hii{} regions as they move into lower density regions of the cloud.
A variety of hybrid models were investigated by \citet{2006ApJS..165..283A} including champagne flow plus stellar wind and a combination of bow shock and champagne flow models in which the density gradient and the strength of the stellar wind were varied.
With the inclusion of a stellar wind, it was discovered that it is possible for a champagne flow to produce limb-brightened morphologies. 
This is due to the fact that the stellar wind creates a dense shell that acts as a barrier to photo-evaporated flows, which divert around it.
\citet{2006ApJS..165..283A} found that for simple bow shock models the line widths are highest ahead of the massive star and for simple champagne flow models the widths are highest in the tail.
They found that for hybrid models, a slow moving star in a steep density gradient has larger velocity widths toward the tail, but a fast moving star in shallow density gradients has its largest velocity widths nearer the star. 
Making sure to remove the effect of any cloud velocity, a hybrid model that matches the line data of a cometary \hii{} region could be used to gather information about the density structure of the natal cloud.

Three-dimensional simulations of \hii{} regions expanding off-centre in turbulent, self-gravitating power-law cores were run by \citet{2007ApJ...668..980M}.
They found that the expanding \hii{} regions were roughly spherical and noted that this is consistent with the analytical results of \citet{1992ApJ...398..184K}.
These results were also confirmed by \citet{2007ApJ...670..471A} who performed similar simulations for $r^{-2}$ and $r^{-3}$ power-law density cores.

New Galactic plane surveys are revisiting the lifetime problem and providing large, well selected samples.
The Galaxy-wide \gls*{rms} survey found \num{\sim 900} mid-\gls*{ir} bright compact \hii{} regions \citep{2013ApJS..208...11L}.
\citet{2011ApJ...730L..33M} used the results of this survey and determined the lifetime of the compact \hii{} phase to be \SI{300}{\kilo\year} or \SIrange{\sim 3}{10}{\percent} of the source's main-sequence lifetime. 
\citet{2011MNRAS.416..972D} simulated the \gls*{rms} results using a particular Galactic gas distribution and different accretion models and compared with the luminosity distribution of the \gls*{rms} survey.
In this work each ionising star in the Galaxy was assumed to be producing \hii{} regions in a uniform density medium and not blowing a stellar wind (i.e. simple Str\"{o}mgren expansion).

The aim of the current paper is to produce a grid of more realistic \gls*{uchii} regions to include in the galaxy model of \citet{2011MNRAS.416..972D} by simulating cometary \hii{} regions.
We explore a parameter space spanning stellar mass, the density of the stellar environment and the age of the star and note the behaviour of some observables across this space.
In a subsequent paper we will use this model grid to provide more realistic \hii{} region sizes and fluxes in an improved galaxy model that will be tested against the \gls*{cornish} survey \citep{2012PASP..124..939H,2013ApJS..205....1P}.
The \gls*{cornish} survey is an arcsecond resolution \SI{5}{\giga\hertz} radio survey of the northern half of the GLIMPSE region ($\SI{10}{\degree} < l < \SI{65}{\degree}$, $|b| < \SI{1}{\degree}$) for compact ionised sources.
With such a large unbiased sample of \gls*{uchii} regions (\num{\sim 240}) it will be possible to test the models in the context of high-mass star formation on the Galactic scale. 

In \cref{sec:model} we introduce the numerical scheme and describe the models we simulated.
The results of the simulations are presented in \cref{sec:results} and we discuss the behaviour of the hot stellar wind region, the ionisation front, the emission measures and the spectral indices when stellar mass, age and cloud density are varied.
We conclude the paper in \cref{sec:conclusions} where we summarise our findings.

\section{The Model}
\label{sec:model}

\subsection{Numerical Scheme}

Our simulations use \gls*{torch}, which is a 3D Eulerian fixed grid fluid dynamics code (code tests are given in \cref{app:code-tests}).
Three coupled physics problems (hydrodynamics, radiative transfer, and heating/cooling) are integrated separately and the result of each is combined to update the solution at each step using the Strang splitting scheme \citep{doi:10.1137/0705041}.

The fluid dynamics was solved on a two-dimensional axisymmetric grid \citep{1991MNRAS.250..581F} using a rotated hybrid HLLC-HLL Riemann solver (see \cref{sec:hll-rot}) to calculate numerical fluxes on each inter-cell boundary.
Left and right Riemann states are interpolated via a MUSCL scheme \citep{1979JCoPh..32..101V} that uses the van Albada slope limiter \citep{1982A&A...108...76V}.
Rotated hybrid Riemann solvers automatically apply fewer wave solvers normal to shocks (the HLL solver in this code) in order to eliminate carbuncle instabilities.
These solvers are very robust and come with an acceptable drop in performance.
The minimum and maximum signal velocities in the HLLC solver are approximated as the minimum and maximum eigenvalues of the Roe matrix \citep{1981JCoPh..43..357R}.
The hydrodynamics scheme is 2nd order accurate in time and space.

The transfer of ionising radiation is calculated using the C\textsuperscript{2}--Ray method by \citet{2006NewA...11..374M}.
At each step the optical depth to each cell is traced via the method of short-characteristics.
In this method the optical depth to the target cell is a weighted average of optical depths in neighbouring cells that lie between the target cell and the cell containing the ionising source.
The ionisation fraction in each cell is implicitly solved before the cell's optical depth can be used in calculations for other cells (which constrains the order in which the cells are iterated over).

Heating from different types of radiation and cooling due to collisional excitation and recombination is calculated using the approximate functions in \citet{2009MNRAS.398..157H}.

The governing hydrodynamic equations are: \hbox{conservation} of mass,
\begin{equation}
\frac{\partial \rho}{\partial t} + \nabla \customdot (\rho \customvec{u}) = \dot{\rho}_\mathrm{w} (\customvec{r}) \mathpunc{;}
\end{equation}
conservation of momentum,
\begin{equation}
\frac{\partial (\rho \customvec{u})}{\partial t} + \nabla \customdot \left(\rho \customvec{u} \otimes \customvec{u}\right) + \nabla p = \customvec{0} \mathpunc{;}
\end{equation}
and conservation of energy,
\begin{equation}
\label{eq:cons-e}
\frac{\partial e}{\partial t} + \nabla \customdot \left(\customvec{u} (e + p)\right) = H - C + \dot{e}_\mathrm{w} (\customvec{r}) \mathpunc{,}
\end{equation}
where $\rho$ is the gas density, $\customvec{u}$ is the fluid velocity, $p$ is the thermal pressure, $e = \frac{1}{\gamma - 1} p + \frac{1}{2} \rho u^2$ is the total energy density with $\gamma = 5 / 3$, $H$ and $C$ are respectively the heating and cooling rates due to atomic/molecular transitions, $\dot{e}_\mathrm{w} (\customvec{r})$ is the injection rate of stellar wind energy density and $\dot{\rho}_\mathrm{w} (\customvec{r})$ is the injection rate of wind material density as a function of position.

We also have equations describing the advection of the hydrogen ionisation fraction,
\begin{equation}
\frac{\partial (f \rho)}{\partial t} + \nabla \customdot \left(f \rho \customvec{u}\right) = \dot{\rho}_\mathrm{w} (\customvec{r}) \mathpunc{,}
\end{equation}
and the rate of hydrogen ionisations and recombinations,
\begin{equation}
\frac{\mathrm{d} f}{\mathrm{d} t} = (1 - f) \left(\Gamma + n_\mathrm{e} C_\mathrm{H}\right) - f n_\mathrm{e} \alpha_\mathrm{H} \mathpunc{,}
\end{equation}
where $f$ is the fraction of hydrogen that is ionised, $n_\mathrm{e}$ is the electron number density, $\Gamma$ is the photo-ionisation rate, $C_\mathrm{H}$ is the collisional ionisation coefficient and $\alpha_\mathrm{H}$ is the recombination coefficient of hydrogen.

Models were simulated on a numerical grid with square cells that have equal side lengths. 
The resolution of each grid is given in \cref{tab:model-params} along with the physical dimensions.
On the edges of the simulation domain we applied Dirichlet boundary conditions. Specifically, we applied reflective conditions on the boundary at $r = 0$ and outflow conditions on the other boundaries.

\subsection{The Star's Environment}

Our model environment has the same density structure as Model F in \citet{2006ApJS..165..283A}.
Stars in these models are off-centre in a spherically symmetric density distribution at a distance of \SI{0.35}{\parsec} from the cloud centre.
The density is given by
\begin{equation}
\label{eq:cloud-density}
	\rho = \rho_0 \left[1 + \left(\frac{r}{r_\mathrm{c}}\right)^2\right]^{\frac{-\alpha}{2}} \mathpunc{,}
\end{equation}
where $r$ is the distance from the cloud centre, $r_\mathrm{c} = \SI{0.01}{\parsec}$ is the cloud core radius, $\rho_0$ is the density at the cloud centre and the power-law index $\alpha$ parameterises the dependence on $r$ when $r \gg r_\mathrm{c}$. 
The density at the star's position is then given by 
\begin{equation}
\label{eq:cloud-density-star}
	\rho_\star = \rho_0 \left[1 + \left(\frac{r_\mathrm{sc}}{r_\mathrm{c}}\right)^2\right]^{\frac{-\alpha}{2}} \mathpunc{,}
\end{equation}
where $r_\mathrm{sc} = \SI{0.35}{\parsec}$ is the distance of the star from the cloud centre, so that $\rho_\star = \frac{1}{1226} \rho_0$.

The power-law index, $\alpha$, has been inferred using a number of different techniques leading to a wide range of values in the literature. 
\citet{2003A&A...409..589H} found indices between \num{1.25} and \num{2.25}.
Five of the seven dark cloud envelopes that were investigated by \citet{1985ApJ...297..436A} were best characterized by an index of $\alpha \simeq 2$.
A range of \numrange{1.0}{1.5} was found by \citet{2000ApJ...537..283V}.
\citet{2009ARep...53.1127P} found an index of $\alpha = 1.6 \pm 0.3$ that falls more steeply in the outer layers of the dense core.
For the current work a value of $\alpha = 2$ was adopted.

The initial pressure in a cloud of constant temperature will have the same structure as the density field.
Without gravity, gas will move down the pressure gradient leading to bulk motion that can interfere with the dynamics of the star's stellar wind and ionisation field, especially over the period of \SI{200}{\kilo\year} each of the simulations here run for.
Instead the pressure of the environment was taken to be uniform such that lower temperatures occur towards the cloud core, and higher temperatures occur away from it (the temperature at the position of the star was \SI{300}{\kelvin}).

\subsection{Parameters}

Using this model environment we explored key parameters, namely the cloud density at the position of the star and the mass of the ZAMS star.
Values for these are given in \cref{tab:model-params}.
In \cref{tab:stellar-params} we give the stellar wind parameters, mass-loss rate and terminal velocity (see \cref{tab:stellar-params}), that were calculated using the predictions of \citet{2001A&A...369..574V}.
These depend on the metallicity (which was assumed to be solar) and also the effective temperature and luminosity of the star. 
We took values for these (for certain stellar masses) from \citet{2011MNRAS.416..972D}, which were calculated using the hydrostatic models of \citet{2000A&A...361..101M}.
For the same masses, Lyman continuum fluxes were also taken from \citet{2011MNRAS.416..972D} who used calculations from \citet{2005A&A...436.1049M} and \citet{2007ApJS..169...83L}.
Effective temperature, luminosity and Lyman continuum flux depend only on stellar mass; hence, this is the only free parameter describing the star.

\begin{table*}
\centering
\caption{The grid of parameters used for the models: stellar mass, $M_\star$; the number density, $\mathrm{n_\star}$, of the cloud of hydrogen gas at the position of the star at the start of the simulation; the number of numerical grid cells along the radial axis, $N_r$; the number of numerical grid cells along the polar axis, $N_z$; and the physical extent of the numerical grid along the radial direction, $L_r$. The other columns are calculated values for each model including: the Str\"{o}mgren radius to star-cloud distance ratio, $y_\mathrm{sc}$; the wind start time, $t_\mathrm{start}$; and the cooling time in the injection region, $t_\mathrm{cool}$.}
\label{tab:model-params}
\sisetup{
table-number-alignment = center,
table-text-alignment = center
}
\begin{tabular}{
S[round-mode=places,round-precision=0,table-format=3.0e0]
S[round-mode=places,round-precision=2,table-format=1.2e1]
S[round-mode=figures,round-precision=3,table-format=3.0e0]
S[round-mode=figures,round-precision=3,table-format=3.0e0]
S[round-mode=places,round-precision=2,table-format=1.2e0]
S[round-mode=places,round-precision=3,table-format=1.3e0]
S[round-mode=places,round-precision=3,table-format=1.3e0]
S[round-mode=places,round-precision=3,table-format=1.3e0]
S[round-mode=places,round-precision=3,table-format=1.3e1]
}
\toprule
{$M_\star$} & {$n_\star$} & {$N_r$} & {$N_z$} & {$L_r$} & {$y_\mathrm{sc}$} & $R_\mathrm{inj}$ & {$t_\mathrm{start}$} & {$t_\mathrm{cool}$} {\TopStrut\BottomStrut} \\
{[\si{\msun}]} & {[\si{\per\centi\meter\cubed}]} & & & {[\si{\parsec}]} & & {[\si{\parsec}]} & {[\si{\kilo\year}]} & {[\si{\kilo\year}]} {\TopStrut\BottomStrut} \\
\midrule

  6 & 8.00e+03 & 150 & 400 & 0.10 & 0.006 & 0.007 & 0.257 & 5.503e+03 \\
  9 & 8.00e+03 & 150 & 300 & 0.15 & 0.018 & 0.010 & 0.357 & 7.468e+04 \\
 12 & 8.00e+03 & 150 & 200 & 0.50 & 0.047 & 0.033 & 1.002 & 9.027e+03 \\
 15 & 8.00e+03 & 150 & 200 & 0.50 & 0.102 & 0.033 & 0.172 & 8.634e+03 \\
 20 & 8.00e+03 & 150 & 200 & 0.50 & 0.214 & 0.033 & 0.026 & 1.703e+04 \\
 30 & 8.00e+03 & 150 & 200 & 1.00 & 0.363 & 0.067 & 0.029 & 1.475e+04 \\
 40 & 8.00e+03 & 150 & 200 & 1.50 & 0.494 & 0.100 & 0.025 & 1.069e+04 \\
 70 & 8.00e+03 & 150 & 200 & 3.00 & 0.681 & 0.200 & 0.035 & 1.086e+04 \\
120 & 8.00e+03 & 150 & 200 & 9.00 & 0.858 & 0.601 & 0.243 & 1.164e+04 \\
  6 & 1.60e+04 & 150 & 300 & 0.06 & 0.004 & 0.004 & 0.111 & 2.577e+03 \\
  9 & 1.60e+04 & 150 & 300 & 0.15 & 0.011 & 0.010 & 0.715 & 2.879e+02 \\
 12 & 1.60e+04 & 150 & 250 & 0.25 & 0.029 & 0.017 & 0.250 & 3.539e+04 \\
 15 & 1.60e+04 & 150 & 200 & 0.50 & 0.064 & 0.033 & 0.344 & 1.425e+04 \\
 20 & 1.60e+04 & 150 & 200 & 0.50 & 0.135 & 0.033 & 0.052 & 1.579e+04 \\
 30 & 1.60e+04 & 150 & 200 & 0.50 & 0.229 & 0.033 & 0.007 & 4.868e+03 \\
 40 & 1.60e+04 & 150 & 200 & 1.00 & 0.311 & 0.067 & 0.015 & 5.620e+03 \\
 70 & 1.60e+04 & 150 & 200 & 2.00 & 0.429 & 0.134 & 0.021 & 5.391e+03 \\
120 & 1.60e+04 & 150 & 200 & 5.00 & 0.540 & 0.334 & 0.083 & 8.106e+03 \\
  6 & 3.20e+04 & 150 & 300 & 0.05 & 0.002 & 0.003 & 0.128 & 1.290e+03 \\
  9 & 3.20e+04 & 150 & 400 & 0.08 & 0.007 & 0.005 & 0.217 & 2.778e+04 \\
 12 & 3.20e+04 & 150 & 300 & 0.20 & 0.019 & 0.013 & 0.256 & 2.828e+04 \\
 15 & 3.20e+04 & 150 & 200 & 0.50 & 0.040 & 0.033 & 0.688 & 3.957e+03 \\
 20 & 3.20e+04 & 150 & 200 & 0.50 & 0.085 & 0.033 & 0.103 & 1.449e+04 \\
 30 & 3.20e+04 & 150 & 200 & 0.50 & 0.144 & 0.033 & 0.015 & 4.869e+03 \\
 40 & 3.20e+04 & 150 & 200 & 1.00 & 0.196 & 0.067 & 0.030 & 1.559e+03 \\
 70 & 3.20e+04 & 150 & 200 & 3.00 & 0.270 & 0.200 & 0.139 & 1.052e+04 \\
120 & 3.20e+04 & 150 & 200 & 5.00 & 0.340 & 0.334 & 0.167 & 2.381e+03 \\
  6 & 6.40e+04 & 150 & 300 & 0.03 & 0.001 & 0.002 & 0.055 & 7.395e+02 \\
  9 & 6.40e+04 & 150 & 300 & 0.05 & 0.004 & 0.003 & 0.106 & 9.792e+02 \\
 12 & 6.40e+04 & 150 & 200 & 0.20 & 0.012 & 0.013 & 0.513 & 1.746e+04 \\
 15 & 6.40e+04 & 150 & 300 & 0.25 & 0.025 & 0.017 & 0.172 & 1.477e+04 \\
 20 & 6.40e+04 & 150 & 200 & 0.50 & 0.053 & 0.033 & 0.207 & 1.139e+04 \\
 30 & 6.40e+04 & 150 & 200 & 0.50 & 0.091 & 0.033 & 0.029 & 4.873e+03 \\
 40 & 6.40e+04 & 150 & 200 & 0.50 & 0.123 & 0.033 & 0.007 & 1.559e+03 \\
 70 & 6.40e+04 & 150 & 200 & 0.50 & 0.170 & 0.033 & 0.001 & 9.800e+02 \\
120 & 6.40e+04 & 150 & 200 & 2.00 & 0.214 & 0.134 & 0.021 & 1.565e+03 \\
  6 & 1.28e+05 & 150 & 300 & 0.03 & 0.001 & 0.002 & 0.064 & 3.734e+02 \\
  9 & 1.28e+05 & 150 & 300 & 0.04 & 0.003 & 0.003 & 0.108 & 7.410e+03 \\
 12 & 1.28e+05 & 150 & 300 & 0.10 & 0.007 & 0.007 & 0.128 & 4.389e+03 \\
 15 & 1.28e+05 & 150 & 300 & 0.20 & 0.016 & 0.013 & 0.176 & 9.584e+03 \\
 20 & 1.28e+05 & 150 & 200 & 0.50 & 0.034 & 0.033 & 0.414 & 5.538e+03 \\
 30 & 1.28e+05 & 150 & 200 & 0.50 & 0.057 & 0.033 & 0.058 & 4.829e+03 \\
 40 & 1.28e+05 & 150 & 200 & 0.50 & 0.078 & 0.033 & 0.015 & 1.559e+03 \\
 70 & 1.28e+05 & 150 & 200 & 0.50 & 0.107 & 0.033 & 0.003 & 3.986e+02 \\
120 & 1.28e+05 & 150 & 200 & 0.50 & 0.135 & 0.033 & 0.001 & 4.178e+02 \\

\bottomrule
\end{tabular}
\end{table*}

\begin{table*}
\centering
\caption{The stellar parameters used for the models. In order, the columns show: stellar mass, $M_\star$; effective temperature, $T_\mathrm{eff}$; stellar radius, $R_\star$; Luminosity, $L_\star$; Lyman continuum flux, $Q_\mathrm{Lyc}$; the mechanical luminosity of the wind, $\mathscr{L}$; mass-loss rate, $\dot{M}$; the free-flowing wind speed, $\mathrm{v_{\infty}}$.}
\label{tab:stellar-params}
\sisetup{
table-number-alignment = center,
table-text-alignment = center
}
\begin{tabular}{
S[round-mode=places,round-precision=0,table-format=3.0e0]
S[round-mode=places,round-precision=0,table-format=5.0e0]
S[round-mode=places,round-precision=1,table-format=2.1e0]
S[round-mode=places,round-precision=2,table-format=1.2e0]
S[round-mode=places,round-precision=2,table-format=2.2e0]
S[round-mode=places,round-precision=2,table-format=1.2e2]
S[round-mode=places,round-precision=2,table-format=1.2e-2]
S[round-mode=places,round-precision=2,table-format=1.2e1]
}
\toprule
{$M_\star$} & {$T_\mathrm{eff}$} & {$R_\star$} & {$\log_{10}(L_\star)$} & {$\log_{10}(Q_\mathrm{Lyc})$} & {$\mathscr{L}$} & {$\dot{M}$} & {$v_{\infty}$} {\TopStrut\BottomStrut} \\
{[\si{\msun}]} & {[\si{\kelvin}]} & {[\si{\rsun}]} & {[\si{\lsun}]} & {[\si{\per\second}]} & {[\si{\erg\per\second}]} & {[\si{\msun\per\year}]} & {[\si{\kilo\meter\per\second}]} {\TopStrut\BottomStrut} \\
\midrule

  6 & 19000 & 3.00 & 3.01 & 43.33 & 5.38e+31 & 1.33e-10 & 1.13e+03 \\
  9 & 22895 & 3.90 & 3.57 & 44.76 & 1.30e+32 & 6.95e-11 & 2.43e+03 \\
 12 & 26743 & 4.50 & 3.96 & 46.02 & 1.72e+33 & 8.03e-10 & 2.60e+03 \\
 15 & 29783 & 5.10 & 4.26 & 47.03 & 1.00e+34 & 4.30e-09 & 2.71e+03 \\
 20 & 33824 & 6.00 & 4.61 & 48.00 & 6.67e+34 & 2.57e-08 & 2.86e+03 \\
 30 & 38670 & 7.30 & 5.02 & 48.69 & 4.73e+35 & 1.54e-07 & 3.11e+03 \\
 40 & 42610 & 8.70 & 5.34 & 49.09 & 1.86e+36 & 5.74e-07 & 3.20e+03 \\
 70 & 47662 & 12.00 & 5.81 & 49.51 & 1.07e+37 & 2.90e-06 & 3.41e+03 \\
120 & 50853 & 17.10 & 6.23 & 49.81 & 4.13e+37 & 1.11e-05 & 3.42e+03 \\

\bottomrule
\end{tabular}
\end{table*}

\subsection{Stellar Winds}
\label{sec:stellar-winds}

\citet{2001A&A...369..574V} determined relations for mass-loss rates of O and B stars for a range of metallicities by using a Monte Carlo radiative transfer method \citep{1999A&A...350..181V, 2000A&A...362..295V} on model atmospheres produced by the Improved Sobolev Approximation code \citep{1993A&A...277..561D, 1997ApJ...477..792D}.
This method allowed them to find the radiative momentum transferred to the wind via photon absorptions and scatterings.
Self-consistent solutions, i.e. models for which the radiative momentum was equal to the wind momentum of the model atmosphere, were then used to find mass-loss rates.
Strictly speaking, the \citet{2000A&A...362..295V} model is applicable in the range $\SI{15}{\msun} \leq M_\star \leq \SI{120}{\msun}$; we are assuming the author's results can be extended to predict stellar wind parameters for our model stars with $M_\star < \SI{15}{\msun}$.

\begin{figure}
	\includegraphics[width=1.00\linewidth]{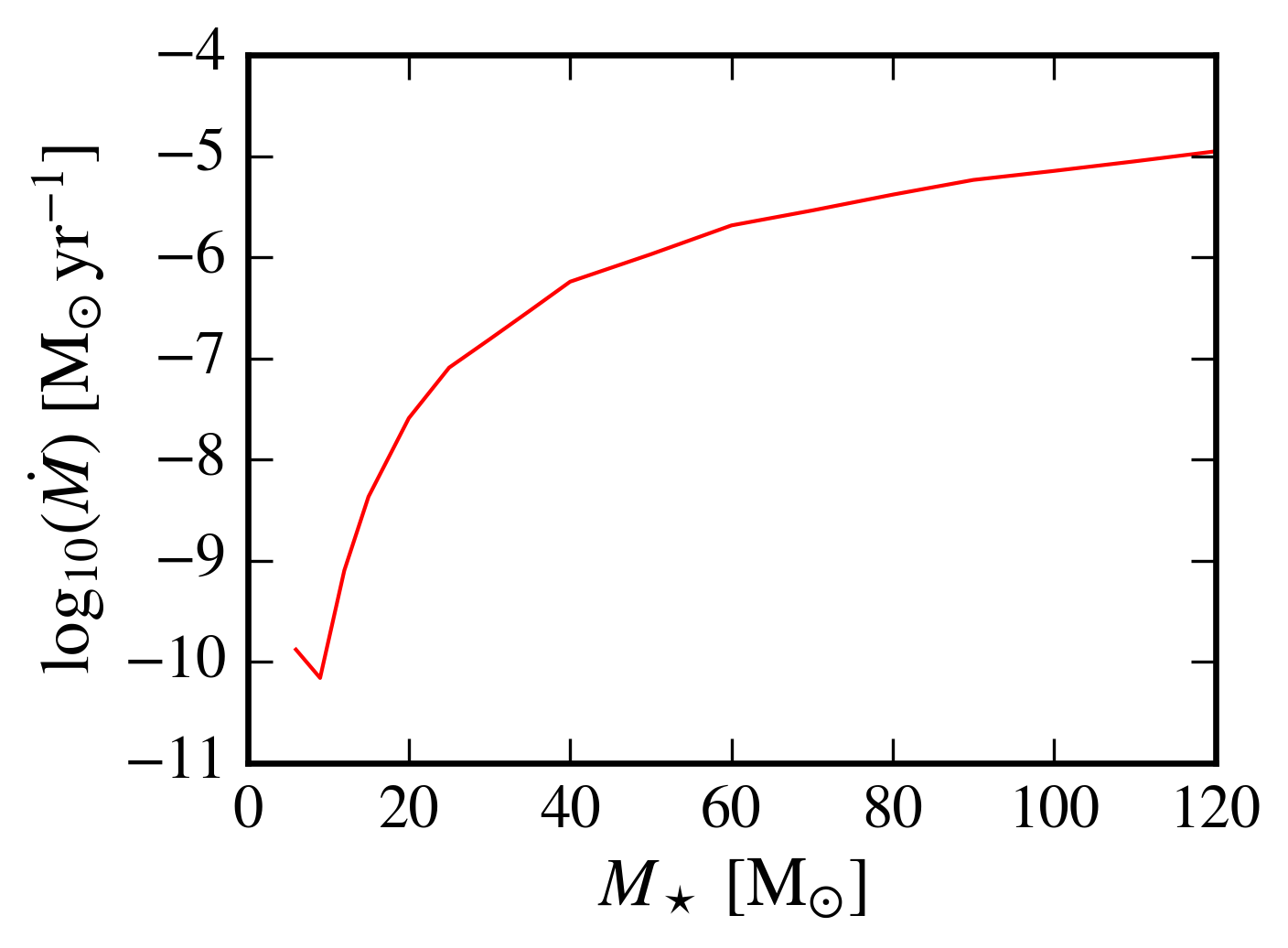}
	\caption{The mass-loss rate of stars plotted against stellar mass.}
	\label{fig:mass-mdot}
\end{figure}

\begin{figure}
	\includegraphics[width=1.00\linewidth]{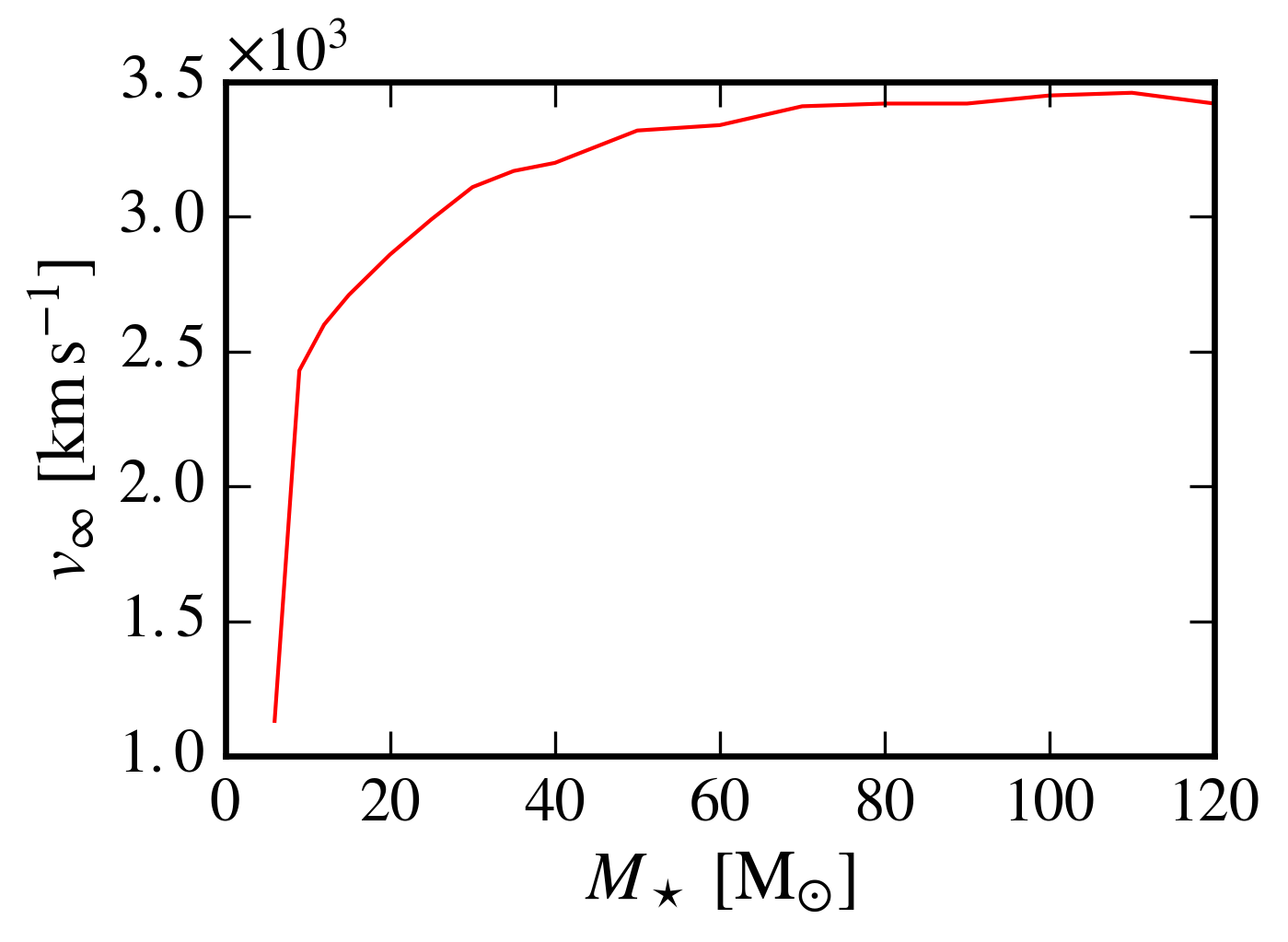}
	\caption{The terminal velocity of stellar winds plotted against the driving star's mass.}
	\label{fig:mass-vinf}
\end{figure}

Mass-loss rates and terminal velocities were calculated for each of our model stars using the values in \cref{tab:stellar-params} and these are plotted in \cref{fig:mass-mdot,fig:mass-vinf}, respectively.
As can be seen in the table the mass injection rate increases for higher mass stars except between \SI{6}{\msun} and \SI{9}{\msun} where there is a decrease.
This is due to the bi-stability jump; the star with $M_\star = \SI{6}{\msun}$ is the only one that lies on the cold side of the bi-stability jump where Fe \upperroman{4} recombines to Fe \upperroman{3}, an efficient line driver.

In order to simulate the effects of a stellar wind we use the thermal energy injection method developed by \citet{1985Natur.317...44C}, which has also been used by \citet{1997A&A...326.1195C} and \citet{2006ApJS..165..283A}.
Due to the resolution of the numerical grid used to simulate our models we were unable to define a free flowing wind region around the star \citep[as in][]{1985A&A...143...59R} that is sufficiently resolved (low resolution regions seed instabilities that grow enough to render the simulation results unphysical).
The stellar wind power is mostly converted to thermal energy at the reverse shock \citep{1975ApJ...200L.107C} so we do not need to reproduce the structure of the unshocked wind region.
Instead we can inject the wind luminosity as a rate of mass, $\dot{M}$, and energy, $\frac{1}{2} \dot{M} v_\infty^2$, into the shocked wind region as it is this power that determines the evolution and structure of the bubble \citep{1977ApJ...218..377W}.
To ensure that the gas in the injection region of our models is adiabatic the gas is not cooled within the injection region in our simulations.

This method also leads to a shocked stellar wind region with the correct temperature.
Inside the injection region the temperature is set by the energy injected per unit mass ($\dot{E} / \dot{M}$) and the average particle mass.
It is nearly uniform in this region, but drops rapidly at its edge as the gas expands supersonically into its surroundings.
The supersonic gas at some point passes through a reverse shock, and is re-thermalised to approximately the same temperature as within the wind injection region.
The post-shock temperature of the gas is insensitive to the exact location of the shock.
If the shock is very near to the edge of the injection region, the pre-shock gas will be only mildly supersonic.
The relatively weak shock and consequently relatively small temperature jump in this case is offset by the higher pre-shock temperature of the gas.
On the other hand, if the shock occurs at a greater distance the pre-shock temperature of the wind may drop substantially lower, but its Mach number rises commensurably, and the kinetic energy per unit mass of this gas asymptotes to the original thermal energy per unit mass (ready to be thermalised again).
If the reverse shock occurs inside the wind injection region the energy and mass within the region builds (keeping the gas at the original temperature) until its pressure overcomes that of the surroundings.

Wind material density and energy density are added within the injection radius at each time-step such that their integrated rates over the volume are $\dot{M}$ and $\frac{1}{2} \dot{M} v_\infty^2$ respectively.
The rate of injected wind energy density is given by $\dot{e}_\mathrm{w} = \frac{1}{2} \dot{\rho}_\mathrm{w} v_\infty^2$, where $\dot{\rho}_\mathrm{w}$ is the rate of injected wind material density: 
\begin{equation}
\dot{\rho}_\mathrm{w} (\customvec{r}) =
\begin{cases}
	\quad \frac{\dot{M}}{\frac{4}{3} \pi R_\mathrm{inj}^3} \mathpunc{,} & \quad \text{if  } |\customvec{r}| < R_\mathrm{inj}  \mathpunc{,} \\
	\quad 0 \mathpunc{,} & \quad \text{otherwise} \mathpunc{,}
\end{cases}
\end{equation}
where $R_\mathrm{inj}$ is the injection radius, which is given in \cref{tab:model-params} for each model.

We tested and confirmed that the evolution of a spherically symmetric wind-blown bubble in a neutral medium with no radiative cooling is as predicted using this method.
The size of the injection region ($\sim 10$ cells) in all our models was chosen to be large enough that the region is well resolved by the numerical grid and small enough that the injection region has little effect on the evolution of the bubble.
An undesirable effect of this is that the injection radius is the radius the wind bubble starts with so that the wind has had a head-start.
It does, however, take time to inject enough pressure in the region to blow a wind.
This is approximately the time it takes to add pressure into the injection region that is comparable to the ambient pressure i.e.
\begin{equation}
t_\mathrm{start} = \frac{p_\mathrm{i}}{\dot{p}} = \frac{n_\mathrm{i} k_\mathrm{B} T_\mathrm{i} (\gamma - 1)}{\dot{e}_\mathrm{w}} \mathpunc{,}
\end{equation}
where $p_\mathrm{i}$, $n_\mathrm{i}$ and $T_\mathrm{i}$ are the pressure, hydrogen number density and temperature of the ionised ambient gas, $k_\mathrm{B}$ is the Boltzmann constant and $\gamma = 5 / 3$ is the ratio of heat capacities.
The start times, $t_\mathrm{start}$, are given in \cref{tab:model-params} for all the models.

\begin{figure}
	\includegraphics[width=1.00\linewidth]{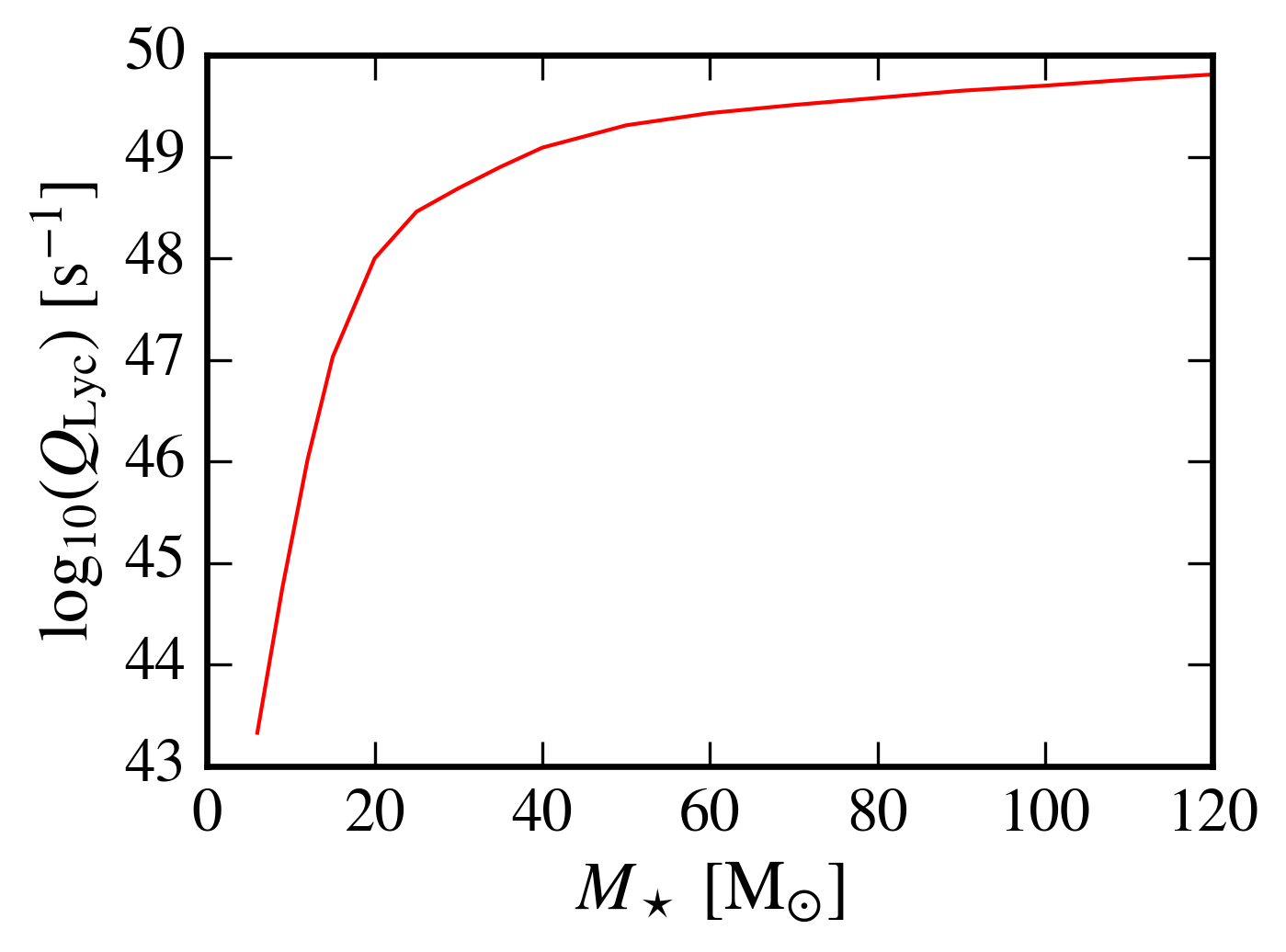}
	\caption{The Lyman continuum photon rate of stars plotted against stellar mass.}
	\label{fig:mass-qlyc}
\end{figure}

\subsection{Radiation Field}

Ionising radiation from each star was assumed to be monochromatic.
Non-ionising \gls*{fuv} radiation was also included to heat the gas in the same way as \citet{2009MNRAS.398..157H}. 
It was assumed that recombinations to the ground state of hydrogen are locally reabsorbed i.e. the diffuse field was treated under the on-the-spot approximation.
The recombination coefficient was therefore calculated using cubic spline interpolation of the case-B recombination coefficient data in \citet{1994MNRAS.268..109H}.
Similarly, the collisional ionisation of neutral hydrogen was calculated by cubic spline interpolation of the data in \citet{1997ApJS..109..517R}.
The cross-section for ionising photons was taken from \citet{1989agna.book.....O}.
\Cref{fig:mass-qlyc} shows the Lyman continuum flux plotted against stellar mass for our models.

\section{Results and Discussion}
\label{sec:results}

\begin{figure*}
	\includegraphics[width=1.00\linewidth]{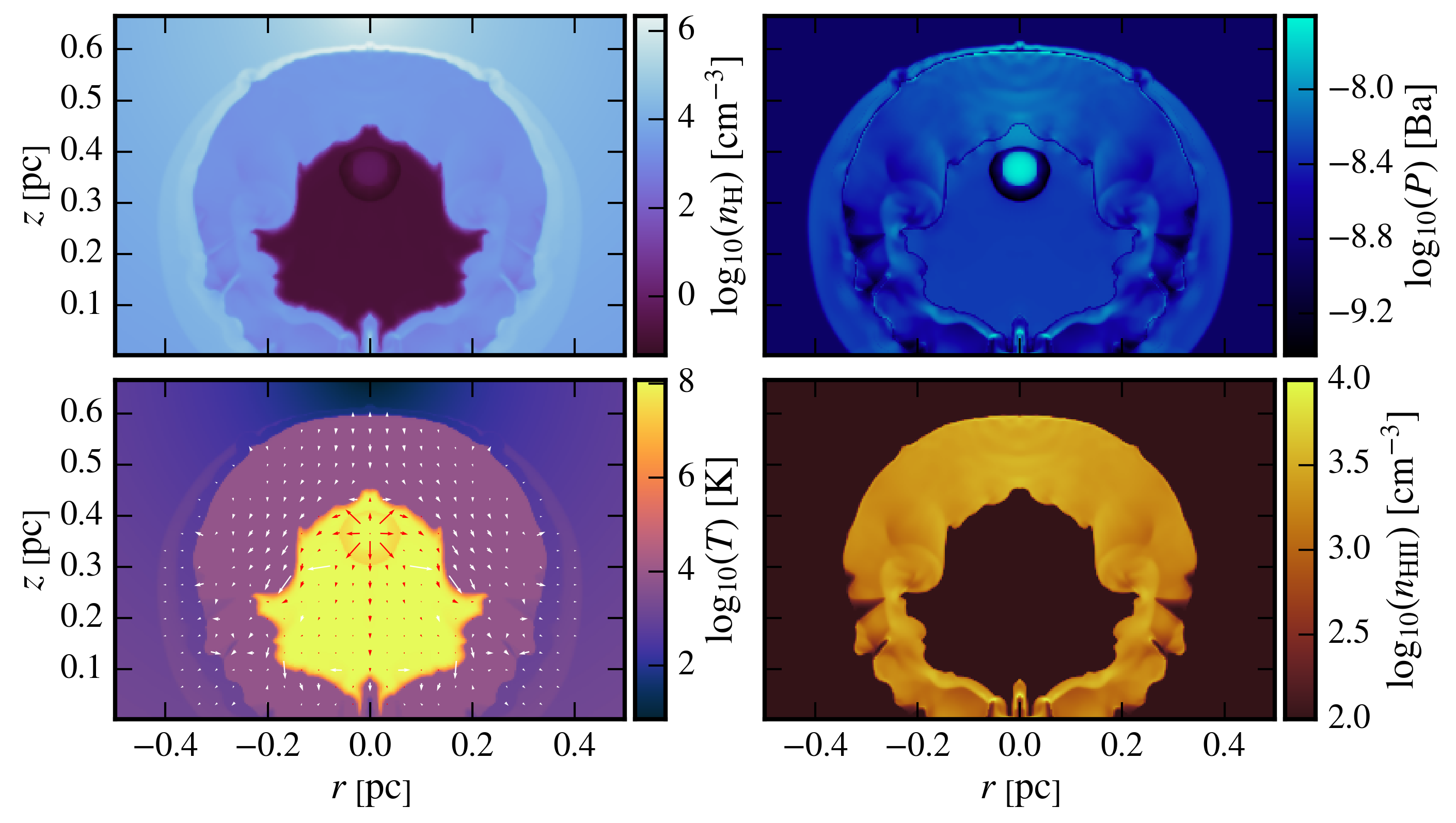}
	\caption{Maps of hydrogen number density (top left), pressure (top right), temperature (bottom left) and ionised hydrogen number density (bottom right) of a model \hii{} region around a star of mass $M_\star = \SI{30}{\msun}$ with a local hydrogen number density of $n_\star = \SI{3.2e4}{\per\centi\meter\cubed}$ at an age of $t = \SI{50}{\kilo\year}$. The star is located at $z = \SI{0.37}{\parsec}$ and $r = \SI{0}{\parsec}$. The blue arrows on the bottom left plot represent velocities of $\SI{3}{\kilo\metre\per\second} < v \leq \SI{30}{\kilo\metre\per\second}$ and the red arrows represent velocities of $\SI{30}{\kilo\metre\per\second} < v \leq \SI{2700}{\kilo\metre\per\second}$. The velocity vectors are separated by ten cells in both dimensions.}
	\label{fig:four-panel}
\end{figure*}

Plots of hydrogen number density, pressure, temperature and ionised hydrogen number density are given in \cref{fig:four-panel} for a typical cometary \hii{} region around a star with mass $M_\star = \SI{30}{\msun}$ at an age of $t = \SI{50}{\kilo\year}$ that starts the simulation in a local hydrogen number density of $n_\star = \SI{3.2e4}{\per\centi\meter\cubed}$.
Nearest the star is the wind injection region of radius \SI{0.04}{\parsec} from which a wind flows outwards becoming supersonic at its edge.
The central temperature is \SI[retain-unity-mantissa=false]{\sim 1e8}{\kelvin} and decreases outside of the injection radius to the wind reverse shock.
Gas outside the injection region is accelerated to supersonic speeds.
We refer to the gas between the wind injection region and the reverse shock as the ``unshocked stellar wind region''.
Downstream of the reverse shock is low density shocked stellar wind material, which is at a uniform temperature of \SI[retain-unity-mantissa=false]{\sim 1e8}{\kelvin} and subsonic.
Further downstream lies the contact discontinuity separating the shocked stellar wind region and the cooler (\SI{\sim 8000}{\kelvin}) ionised ambient gas.
Earlier on in the evolution of this \hii{} region we would expect to see a shock ahead of the contact discontinuity, but pressures either side of the shock have equilibrated so that the shock no longer exists.
Finally, the ionised region is bounded by an ionisation front that separates the ionised gas from the neutral gas shock.
Photo-ionisation has heated the gas in the ionised region and therefore has created an over-pressure with respect to the surrounding ambient gas, which explains the presence of the shock just outside of the ionisation front.

\subsection{Ionisation Fronts}

\begin{table}
\centering
\caption{Measurements of $D_\mathrm{i}(\theta)$, which is the distance from the star to the edge of the ionised region along a direction with polar angle $\theta$. The analytical ionisation front radius, $R_\mathrm{raga1}$ is also given by numerically solving \cref{equ:raga-1}. These measurements were made for stars of age of \SI{50}{\kilo\year}.}
\label{tab:sizes-mvd-if}
\sisetup{
table-number-alignment=center,
table-text-alignment=center,
table-format=1.4e0
}
\begin{tabular}{
S[round-mode=places,round-precision=0,table-format=3.0e0]
S[table-text-alignment=left]
S[round-mode=figures,round-precision=2,table-format=1.3e0]
S[round-mode=figures,round-precision=2,table-format=1.3e0]
S[round-mode=figures,round-precision=2,table-format=1.3e0]
S[round-mode=figures,round-precision=2,table-format=1.4e0]
S[round-mode=figures,round-precision=2,table-format=1.4e0]
}
\toprule
{$M_\star$} & \multicolumn{1}{c}{Dist.} & \multicolumn{5}{c}{$\mathrm{n_\star}$ [\SI[retain-unity-mantissa=false]{1e4}{\per\centi\meter\cubed}]} \TopStrut{}\BottomStrut{} \\
\cline{3-7}
\multicolumn{1}{c}{[\si{\msun}]} & \multicolumn{1}{c}{[\si{\parsec}]} & {\num[round-mode=places,round-precision=1]{0.8}} & {\num[round-mode=places,round-precision=1]{1.6}} & {\num[round-mode=places,round-precision=1]{3.2}} & {\num[round-mode=places,round-precision=1]{6.4}} & {\num[round-mode=places,round-precision=1]{12.8}} \TopStrut{}\BottomStrut \\
\midrule

{\num{6.0}} & {$D_\mathrm{i}(0)$} & 0.0338235294118 & 0.0185294117647 & 0.01 & 0.00970588235294 & 0.00558823529412 \\
& {$D_\mathrm{i}(\frac{\pi}{2})$} & 0.0367647058824 & 0.0229411764706 & 0.0135294117647 & 0.0103676470588 & 0.00617647058824 \\
& {$D_\mathrm{i}(\pi)$} & 0.0426470588235 & 0.0255882352941 & 0.0158823529412 & 0.0105882352941 & 0.00764705882353 \\
& {$R_\mathrm{raga1}$} & 0.0306859707059 & 0.0195131462556 & 0.0123048025217 & 0.00775174056656 & 0.00488329091732 \\
& & & & & & \\
{\num{9.0}} & {$D_\mathrm{i}(0)$} & 0.0776470588235 & 0.0529411764706 & 0.0388235294118 & 0.0211764705882 & 0.0139705882353 \\
& {$D_\mathrm{i}(\frac{\pi}{2})$} & 0.0847058823529 & 0.0595588235294 & 0.0388235294118 & 0.0233823529412 & 0.0143382352941 \\
& {$D_\mathrm{i}(\pi)$} & 0.105882352941 & 0.0705882352941 & 0.0420588235294 & 0.0247058823529 & 0.0154411764706 \\
& {$R_\mathrm{raga1}$} & 0.0788303994758 & 0.0545942960557 & 0.0361606198996 & 0.0231665456399 & 0.0146326967894 \\
& & & & & & \\
{\num{12.0}} & {$D_\mathrm{i}(0)$} & 0.148823529412 & 0.095 & 0.0772058823529 & 0.0617647058824 & 0.035 \\
& {$D_\mathrm{i}(\frac{\pi}{2})$} & 0.169117647059 & 0.122941176471 & 0.0882352941176 & 0.0617647058824 & 0.0401470588235 \\
& {$D_\mathrm{i}(\pi)$} & 0.189411764706 & 0.150882352941 & 0.110294117647 & 0.0705882352941 & 0.0452941176471 \\
& {$R_\mathrm{raga1}$} & 0.149418406251 & 0.112147654271 & 0.0813785954988 & 0.0565973539386 & 0.0376330125188 \\
& & & & & & \\
{\num{15.0}} & {$D_\mathrm{i}(0)$} & 0.195588235294 & 0.158823529412 & 0.138970588235 & 0.0926470588235 & 0.0698529411765 \\
& {$D_\mathrm{i}(\frac{\pi}{2})$} & 0.257352941176 & 0.202941176471 & 0.152205882353 & 0.110661764706 & 0.0790441176471 \\
& {$D_\mathrm{i}(\pi)$} & 0.308823529412 & 0.238235294118 & 0.178676470588 & 0.133823529412 & 0.0955882352941 \\
& {$R_\mathrm{raga1}$} & 0.230060688129 & 0.178914084899 & 0.136636237405 & 0.101552157209 & 0.0727458585652 \\
& & & & & & \\
{\num{20.0}} & {$D_\mathrm{i}(0)$} & 0.264705882353 & 0.235294117647 & 0.191176470588 & 0.169852941176 & 0.132352941176 \\
& {$D_\mathrm{i}(\frac{\pi}{2})$} & 0.397058823529 & 0.317647058824 & 0.24375 & 0.190073529412 & 0.145588235294 \\
& {$D_\mathrm{i}(\pi)$} & 0.476470588235 & 0.376470588235 & 0.296323529412 & 0.218382352941 & 0.165441176471 \\
& {$R_\mathrm{raga1}$} & 0.338464834648 & 0.266780701564 & 0.208972823105 & 0.16153111741 & 0.122203703268 \\
& & & & & & \\
{\num{30.0}} & {$D_\mathrm{i}(0)$} & 0.296470588235 & 0.264705882353 & 0.225 & 0.211764705882 & 0.176470588235 \\
& {$D_\mathrm{i}(\frac{\pi}{2})$} & 0.555882352941 & 0.430147058824 & 0.344117647059 & 0.264705882353 & 0.207352941176 \\
& {$D_\mathrm{i}(\pi)$} & 0.778235294118 & 0.582352941176 & 0.423529411765 & 0.341176470588 & 0.238235294118 \\
& {$R_\mathrm{raga1}$} & 0.444866959306 & 0.350390656221 & 0.276281696291 & 0.216702665299 & 0.167912095465 \\
& & & & & & \\
{\num{40.0}} & {$D_\mathrm{i}(0)$} & 0.305294117647 & 0.298823529412 & 0.264705882353 & 0.235294117647 & 0.2 \\
& {$D_\mathrm{i}(\frac{\pi}{2})$} & 0.737794117647 & 0.560294117647 & 0.441176470588 & 0.338235294118 & 0.264705882353 \\
& {$D_\mathrm{i}(\pi)$} & 1.06852941176 & 0.747058823529 & 0.558823529412 & 0.470588235294 & 0.329411764706 \\
& {$R_\mathrm{raga1}$} & 0.523264931995 & 0.410405158844 & 0.323592755874 & 0.25488580812 & 0.199267110646 \\
& & & & & & \\
{\num{70.0}} & {$D_\mathrm{i}(0)$} & 0.334852941176 & 0.323529411765 & 0.33 & 0.264705882353 & 0.235294117647 \\
& {$D_\mathrm{i}(\frac{\pi}{2})$} & 1.19058823529 & 0.867647058824 & 0.77 & 0.476470588235 & 0.389705882353 \\
& {$D_\mathrm{i}(\pi)$} & 1.37661764706 & 1.02941176471 & 0.88 & 0.6 & 0.470588235294 \\
& {$R_\mathrm{raga1}$} & 0.623984779848 & 0.485745501339 & 0.38184324724 & 0.30118010541 & 0.236862043722 \\
& & & & & & \\
{\num{120.0}} & {$D_\mathrm{i}(0)$} & 1.3325 & 0.391764705882 & 0.392647058824 & 0.319264705882 & 0.280147058824 \\
& {$D_\mathrm{i}(\frac{\pi}{2})$} & 1.95955882353 & 1.24875 & 1.13867647059 & 0.712205882353 & 0.541617647059 \\
& {$D_\mathrm{i}(\pi)$} & 2.35147058824 & 1.56705882353 & 1.335 & 0.884117647059 & 0.747058823529 \\
& {$R_\mathrm{raga1}$} & 0.710951713132 & 0.549411621369 & 0.430144158585 & 0.338974269898 & 0.267187236491 \\

\bottomrule
\end{tabular}
\end{table}

\begin{table}
\centering
\caption{Measurements of $D_\mathrm{i}(\theta)$, which is the distance from the star to the edge of the ionised region along a direction with polar angle $\theta$. The analytical ionisation front radius, $R_\mathrm{raga1}$ is also given by numerically solving \cref{equ:raga-1}. These measurements were made for stars in a local number density of \SI{3.2e4}{\per\centi\meter\cubed}.}
\label{tab:sizes-mvt-if}
\sisetup{
table-number-alignment = center,
table-text-alignment = center,
table-format=1.3e0
}
\begin{tabular}{
S[round-mode=places,round-precision=0,table-format=3.0e0]
S[table-text-alignment=left]
S[round-mode=figures,round-precision=2,table-format=1.3e0]
S[round-mode=figures,round-precision=2,table-format=1.3e0]
S[round-mode=figures,round-precision=2,table-format=1.3e0]
S[round-mode=figures,round-precision=2,table-format=1.3e0]
S[round-mode=figures,round-precision=2,table-format=1.3e0]
}
\toprule
{$M_\star$} & \multicolumn{1}{c}{Dist.} & \multicolumn{5}{c}{Age [\si{\kilo\year}]} \TopStrut{}\BottomStrut{} \\
\cline{3-7}
\multicolumn{1}{c}{[\si{\msun}]} & \multicolumn{1}{c}{[\si{\parsec}]} & {\num[round-mode=figures,round-precision=1]{20}} & {\num[round-mode=figures,round-precision=1]{40}} & {\num[round-mode=figures,round-precision=1]{60}} & {\num[round-mode=figures,round-precision=1]{80}} & {\num[round-mode=figures,round-precision=1]{100}} \TopStrut{}\BottomStrut{} \\
\midrule

{\num{6.0}} & {$D_\mathrm{i}(0)$} & 0.0147058823529 & 0.0123529411765 & 0.0117647058824 & 0.0158823529412 & 0.0141176470588 \\
& {$D_\mathrm{i}(\frac{\pi}{2})$} & 0.0158823529412 & 0.0144117647059 & 0.015 & 0.0164705882353 & 0.0161764705882 \\
& {$D_\mathrm{i}(\pi)$} & 0.0188235294118 & 0.0149264705882 & 0.0194117647059 & 0.0194117647059 & 0.0182352941176 \\
& {$R_\mathrm{raga1}$} & 0.0121816622457 & 0.012302789183 & 0.0123050784501 & 0.0123051212377 & 0.0123051220246 \\
& & & & & & \\
{\num{9.0}} & {$D_\mathrm{i}(0)$} & 0.0382352941176 & 0.0410294117647 & 0.0411764705882 & 0.0376470588235 & 0.0382352941176 \\
& {$D_\mathrm{i}(\frac{\pi}{2})$} & 0.0360294117647 & 0.0410294117647 & 0.0404411764706 & 0.0388235294118 & 0.0382352941176 \\
& {$D_\mathrm{i}(\pi)$} & 0.0411764705882 & 0.0436764705882 & 0.0426470588235 & 0.0388235294118 & 0.05 \\
& {$R_\mathrm{raga1}$} & 0.0313465178801 & 0.0354790826332 & 0.0365082915083 & 0.0367787697763 & 0.0368506687268 \\
& & & & & & \\
{\num{12.0}} & {$D_\mathrm{i}(0)$} & 0.0588235294118 & 0.0776470588235 & 0.0847058823529 & 0.0919117647059 & 0.0955882352941 \\
& {$D_\mathrm{i}(\frac{\pi}{2})$} & 0.0691176470588 & 0.0873529411765 & 0.0970588235294 & 0.102941176471 & 0.106617647059 \\
& {$D_\mathrm{i}(\pi)$} & 0.0794117647059 & 0.106764705882 & 0.116470588235 & 0.121323529412 & 0.125 \\
& {$R_\mathrm{raga1}$} & 0.0594883327413 & 0.0763756903709 & 0.0850785852748 & 0.0899589963937 & 0.0927952295105 \\
& & & & & & \\
{\num{15.0}} & {$D_\mathrm{i}(0)$} & 0.107058823529 & 0.131764705882 & 0.152941176471 & 0.169411764706 & 0.182647058824 \\
& {$D_\mathrm{i}(\frac{\pi}{2})$} & 0.111176470588 & 0.148235294118 & 0.170588235294 & 0.190588235294 & 0.206470588235 \\
& {$D_\mathrm{i}(\pi)$} & 0.115294117647 & 0.16 & 0.194117647059 & 0.225882352941 & 0.246176470588 \\
& {$R_\mathrm{raga1}$} & 0.0916391717113 & 0.124856290917 & 0.146400370214 & 0.161613188875 & 0.172793998721 \\
& & & & & & \\
{\num{20.0}} & {$D_\mathrm{i}(0)$} & 0.129411764706 & 0.176470588235 & 0.210294117647 & 0.239411764706 & 0.258823529412 \\
& {$D_\mathrm{i}(\frac{\pi}{2})$} & 0.152941176471 & 0.220588235294 & 0.262867647059 & 0.304705882353 & 0.329411764706 \\
& {$D_\mathrm{i}(\pi)$} & 0.170588235294 & 0.242647058824 & 0.305882352941 & 0.348235294118 & 0.4 \\
& {$R_\mathrm{raga1}$} & 0.134838025914 & 0.188329868006 & 0.226997490053 & 0.257272123145 & 0.281912171586 \\
& & & & & & \\
{\num{30.0}} & {$D_\mathrm{i}(0)$} & 0.168235294118 & 0.211764705882 & 0.238235294118 & 0.264705882353 & 0.294117647059 \\
& {$D_\mathrm{i}(\frac{\pi}{2})$} & 0.206470588235 & 0.3 & 0.363970588235 & 0.426470588235 & 0.470588235294 \\
& {$D_\mathrm{i}(\pi)$} & 0.244705882353 & 0.376470588235 & 0.45 & 0.5 & 0.632352941176 \\
& {$R_\mathrm{raga1}$} & 0.177213691236 & 0.247964229305 & 0.301540026304 & 0.345257120608 & 0.382269188886 \\
& & & & & & \\
{\num{40.0}} & {$D_\mathrm{i}(0)$} & 0.185294117647 & 0.235294117647 & 0.264705882353 & 0.286764705882 & 0.305882352941 \\
& {$D_\mathrm{i}(\frac{\pi}{2})$} & 0.257352941176 & 0.242647058824 & 0.470588235294 & 0.544852941176 & 0.611764705882 \\
& {$D_\mathrm{i}(\pi)$} & 0.308823529412 & 0.470588235294 & 0.588235294118 & 0.726470588235 & 0.823529411765 \\
& {$R_\mathrm{raga1}$} & 0.208421179515 & 0.29027968707 & 0.353580354597 & 0.406136829711 & 0.451350215327 \\
& & & & & & \\
{\num{70.0}} & {$D_\mathrm{i}(0)$} & 0.235294117647 & 0.335294117647 & 0.291176470588 & 0.338235294118 & 0.344117647059 \\
& {$D_\mathrm{i}(\frac{\pi}{2})$} & 0.360294117647 & 0.670588235294 & 0.808823529412 & 0.997794117647 & 1.14705882353 \\
& {$D_\mathrm{i}(\pi)$} & 0.5 & 0.698529411765 & 0.873529411765 & 0.947058823529 & 1.07058823529 \\
& {$R_\mathrm{raga1}$} & 0.248497349139 & 0.34284422944 & 0.417234938834 & 0.479932282181 & 0.534593890804 \\
& & & & & & \\
{\num{120.0}} & {$D_\mathrm{i}(0)$} & 0.323529411765 & 0.404411764706 & 0.441176470588 & 0.423529411765 & 0.476470588235 \\
& {$D_\mathrm{i}(\frac{\pi}{2})$} & 0.695588235294 & 0.955882352941 & 1.125 & 1.27058823529 & 1.45588235294 \\
& {$D_\mathrm{i}(\pi)$} & 0.808823529412 & 1.25 & 1.32352941176 & 1.50588235294 & 1.58823529412 \\
& {$R_\mathrm{raga1}$} & 0.283088058664 & 0.386813378747 & 0.469671902196 & 0.540163892328 & 0.602115300635 \\

\bottomrule
\end{tabular}
\end{table}

Assuming balance between photo-ionisations and recombinations, for a star in a uniform medium with no stellar wind we have:
\begin{equation}
\label{equ:ion-balance}
Q_\mathrm{Lyc} = \frac{4\pi}{3} n_\mathrm{i}^2(t) \alpha_\mathrm{B} R_\mathrm{IF}^3(t) \mathpunc{,}
\end{equation}
where $n_\mathrm{i}$ is the hydrogen number density inside the ionised region, $\alpha_\mathrm{B}$ is the case-B recombination coefficient for hydrogen and $R_\mathrm{IF}(t)$ is the time-dependent radius of the ionised region.
The Str\"{o}mgren radius is the initial radius of the ionised region:
\begin{equation}
\label{equ:stromgren}
	R_\mathrm{st} = 
	\left(\frac{3 Q_\mathrm{Lyc}}{4 \pi n_{\mathrm H}^2 \alpha_{\mathrm B}}\right)^{1/3} \mathpunc{,}
\end{equation}
where $n_\mathrm{H}$ is the initial hydrogen number density of the uniform medium.
Ionisation heats the gas and sets up an over-pressure; expansion thereafter is driven by this over-pressure.

\citet{2012MNRAS.419L..39R} show that if the shock driven by the over-pressure is isothermal and it is assumed that the expanding \hii{} region is in pressure balance with the shocked neutral gas, the equation of motion of the ionisation front is
\begin{equation}
\label{equ:raga-1}
	\frac{1}{c_\mathrm{i}} \frac{d R_\mathrm{raga1}(t)}{dt} = \left(\frac{R_\mathrm{st}}{R_\mathrm{raga1}(t)}\right)^{3/4} - \frac{c_\mathrm{a}^2}{c_\mathrm{i}^2} \left(\frac{R_\mathrm{st}}{R_\mathrm{raga1}(t)}\right)^{-3/4} \mathpunc{.}
\end{equation}
The Spitzer approximation \citep{1978ppim.book.....S} can be reproduced by this equation by noting that the second term on the right-hand side of equation is negligible at early times:
\begin{equation}
	\label{equ:spitzer}
	R_\mathrm{spitzer} = R_\mathrm{st} \left(1 + \frac{7}{4} \frac{t}{t_\mathrm{s}} \right)^{4/7} \mathpunc{,}
\end{equation}
where $t_\mathrm{s} = R_\mathrm{st} / c_\mathrm{i}$ is the sound crossing time-scale and $c_\mathrm{i}$ is the sound speed of the ionised gas.

According to \citet{2012RMxAA..48..149R} \cref{equ:raga-1} neglects the inertia of the expanding neutral material and therefore derived solutions should under-estimate the distance to the ionisation front. 
Including the inertia gives the equation of the shell,
\begin{equation}
\label{equ:raga-2}
	\frac{d^2 R_\mathrm{raga2}(t)}{d t^2} + \frac{3}{R_\mathrm{raga2}} \left(\frac{d R_\mathrm{raga2}(t)}{d t}\right)^2 = \frac{3R_\mathrm{st}^{3/2} c_\mathrm{i}^2}{R_\mathrm{raga2}^{5/2}} - \frac{3c_\mathrm{a}^2}{R_\mathrm{raga2}} \mathpunc{.}
\end{equation}
At early times the solution can be approximated \citep{2006ApJ...646..240H} by
\begin{equation}
	\label{equ:hosokawa-if}
	R_\mathrm{hi} = R_\mathrm{st} \left(1 + \frac{7}{4}\sqrt{\frac{4}{3}} \frac{t}{t_\mathrm{s}} \right)^{4/7} \mathpunc{.}
\end{equation}
Simulations of D-type expansion of ionisation fronts \citep{2015MNRAS.453.1324B} have shown close agreement with \cref{equ:hosokawa-if} at early times and with \cref{equ:raga-1} at late times.

The sizes of the ionised regions simulated in the current paper are shown in \cref{tab:sizes-mvd-if,tab:sizes-mvt-if}, along with the analytical radius of ionised regions that would evolve around stars with the same stellar parameters but in a uniform density medium (\cref{equ:raga-1}).
The isothermal sound speed in the ionised ambient gas was assumed to be $c_\mathrm{i} = \sqrt{R T_\mathrm{i} / \mu_\mathrm{H}}$, where the ionised gas temperature is $T_\mathrm{i} \simeq \SI{8000}{\kelvin}$, the average molar mass of ionised hydrogen is $\mu_\mathrm{H} = \SI{0.5}{\gram\per\mol}$ and $R$ is the gas constant, giving a sound speed of $c_\mathrm{i} = \SI{11.5}{\kilo\meter\per\second}$.
In \cref{tab:sizes-mvd-if} the measured ionisation regions show a decrease in size for higher local densities as expected.
The sizes also increase for higher masses due to higher Lyman continuum photon fluxes.
There is generally close agreement with the analytical radius, with the simulated regions consistently slightly larger.
This is due to the contribution of the wind to driving the shock ahead of the ionisation front.
Simulations of stars without stellar winds confirm this as they produce \hii{} regions with radii that are very close to those predicted by \cref{equ:raga-1}.
For higher-mass stars the radius of the simulated ionised region is noticeably larger than the analytical radius, suggesting that the strong winds from these stars are contributing relatively more to the expansion of the \hii{} region.

As a \hii{} region expands the pressure inside the region drops until it is equal to the ambient gas pressure, at which point it stagnates. 
The stagnation radius can be found, by setting $d R_\mathrm{raga1}(t) / dt = 0$ in \cref{equ:raga-1}, to be
\begin{equation}
\label{equ:stagnation}
R_\mathrm{stag} = \left(\frac{c_\mathrm{i}}{c_\mathrm{a}}\right)^{4/3} R_\mathrm{st} \mathpunc{,}
\end{equation}
where $c_\mathrm{a}$ is the isothermal sound speed of the neutral ambient gas.
The temperature of the neutral ambient gas is $T_\mathrm{a} \simeq \SI{300}{\kelvin}$, so the sound speed is $c_\mathrm{a} \simeq \SI{1.58}{\kilo\meter\per\second}$.
In our models we therefore have $R_\mathrm{stag} \simeq 14.2 R_\mathrm{st}$.
Setting $R_\mathrm{spitzer}(t_\mathrm{stag}) = R_\mathrm{stag}$ in \cref{equ:spitzer} we can get an approximation for the stagnation time in terms of the sound-crossing time of the initial Str\"{o}mgren sphere: $t_\mathrm{stag} \simeq \num{60} t_\mathrm{s}$.
We can see that for smaller Str\"{o}mgren radii the stagnation time occurs earlier.
\Cref{tab:sizes-mvt-if} shows stagnating ionisation fronts that occur at earlier ages for smaller-mass stars.

Most of the simulations of our models did not show the unbounded ionisation front expansion predicted by \citet{1990ApJ...349..126F} for $\alpha > 3/2$.
Only for stars with masses $M_\star \geq \SI{40}{\msun}$ and local densities $n_\star \leq \SI{1.6e4}{\per\centi\metre\cubed}$ did the ionisation front break free of the cloud and almost always in finger-like structures \citep[a.k.a. the shadowing instability, see][]{1999MNRAS.310..789W}.
In all other models the regions are bounded approximately by a sphere, which is consistent with the three-dimensional simulations of off-centre \gls*{uchii} regions by \citet{2007ApJ...668..980M}.
The regions are at a roughly constant density throughout the simulation and grow to a maximum radius that coincides with when the pressures either side of the ionisation front have equalised.

These results very closely match those in \citet{2007ApJ...670..471A}, in which the condition necessary for \hii{} regions in power-law density environments to remain bounded during their initial formation stage was given by:
\begin{equation}
\label{eq:arthur-bounded}
\frac{1}{3} y_\mathrm{sc}^3 < \frac{2}{(2\alpha - 1)(2\alpha - 2)(2\alpha - 3)} \mathpunc{,}
\end{equation}
where $y_\mathrm{sc} = R_\mathrm{st} / r_\mathrm{sc}$ i.e. the ratio of the Str\"{o}mgren radius for a star in a uniform medium to the distance between the star and the cloud centre.
Values of this ratio for each set of model parameters were calculated and are given in \cref{tab:model-params} by setting $n_\mathrm{H}$ in \cref{equ:stromgren} equal to the local hydrogen number density, $n_\star$.
Our power-law density environments have $\alpha = 2$, so \cref{eq:arthur-bounded} reduces to $y_\mathrm{sc} < 1$, which is true for all our models.
\citet{2007ApJ...670..471A} also found that in $\alpha = 2$ power-law environments, if $y_\mathrm{sc} \lesssim 0.02$ then pressure balance can halt the breakout of the \hii{} region during the expansion stage.
As mentioned before the pressures do equalise before the \hii{} regions can become unbounded for most models.
However, this is also seen for models with $y_\mathrm{sc} > 0.02$.

\subsection{Shocked Stellar Wind Region}

\begin{table}
\centering
\caption{Measurements of $D_\mathrm{s}(\theta)$, which is the distance from the star to the edge of the shocked stellar wind region along a direction with polar angle $\theta$. The analytical radius for a radiative bubble at the pressure confinement time, $R_\mathrm{P}$ (see \cref{equ:confine-radius} and \cref{equ:radius-rb}), is also given. These measurements were made for stars of age \SI{50}{\kilo\year}.}
\label{tab:sizes-mvd-shock}
\sisetup{
table-number-alignment=center,
table-text-alignment=center,
table-format=1.4e0
}
\begin{tabular}{
S[round-mode=places,round-precision=0,table-format=3.0e0]
S[table-text-alignment=left]
S[round-mode=figures,round-precision=2,table-format=1.4e0]
S[round-mode=figures,round-precision=2,table-format=1.4e0]
S[round-mode=figures,round-precision=2,table-format=1.4e0]
S[round-mode=figures,round-precision=2,table-format=1.4e0]
S[round-mode=figures,round-precision=2,table-format=1.4e0]
}
\toprule
{$M_\star$} & \multicolumn{1}{c}{Dist.} & \multicolumn{5}{c}{$\mathrm{n_\star}$ [\SI[retain-unity-mantissa=false]{1e4}{\per\centi\meter\cubed}]} \TopStrut{}\BottomStrut{} \\
\cline{3-7}
\multicolumn{1}{c}{[\si{\msun}]} & \multicolumn{1}{c}{[\si{\parsec}]} & {\num[round-mode=places,round-precision=1]{0.8}} & {\num[round-mode=places,round-precision=1]{1.6}} & {\num[round-mode=places,round-precision=1]{3.2}} & {\num[round-mode=places,round-precision=1]{6.4}} & {\num[round-mode=places,round-precision=1]{12.8}} \TopStrut{}\BottomStrut{} \\
\midrule

{\num{6.0}} & {$D_\mathrm{s}(0)$} & 0.0176470588235 & 0.00970588235294 & 0.00588235294118 & 0.00617647058824 & 0.00352941176471 \\
& {$D_\mathrm{s}(\frac{\pi}{2})$} & 0.0242647058824 & 0.0172058823529 & 0.00882352941176 & 0.00727941176471 & 0.00426470588235 \\
& {$D_\mathrm{s}(\pi)$} & 0.0323529411765 & 0.0176470588235 & 0.0117647058824 & 0.00926470588235 & 0.00617647058824 \\
& {$R_\mathrm{P}$} & 0.00706067494703 & 0.00495717967716 & 0.00333604639872 & 0.00273232936452 & 0.00185279845369 \\
& & & & & & \\
{\num{9.0}} & {$D_\mathrm{s}(0)$} & 0.0158823529412 & 0.0242647058824 & 0.00970588235294 & 0.00352941176471 & 0.00294117647059 \\
& {$D_\mathrm{s}(\frac{\pi}{2})$} & 0.015 & 0.0181985294118 & 0.0169852941176 & 0.00617647058824 & 0.00367647058824 \\
& {$D_\mathrm{s}(\pi)$} & 0.0847058823529 & 0.0176470588235 & 0.0226470588235 & 0.0141176470588 & 0.00661764705882 \\
& {$R_\mathrm{P}$} & 0.00615226789831 & 0.00472399732748 & 0.00342705694604 & 0.00234296725543 & 0.00162357306436 \\
& & & & & & \\
{\num{12.0}} & {$D_\mathrm{s}(0)$} & 0.0405882352941 & 0.0111764705882 & 0.0220588235294 & 0.0205882352941 & 0.0123529411765 \\
& {$D_\mathrm{s}(\frac{\pi}{2})$} & 0.0372058823529 & 0.0223529411765 & 0.0248161764706 & 0.0220588235294 & 0.0175 \\
& {$D_\mathrm{s}(\pi)$} & 0.142058823529 & 0.111764705882 & 0.0625 & 0.0382352941176 & 0.0205882352941 \\
& {$R_\mathrm{P}$} & 0.0175818629217 & 0.0138419162597 & 0.0107933173667 & 0.00825997013842 & 0.00597948106192 \\
& & & & & & \\
{\num{15.0}} & {$D_\mathrm{s}(0)$} & 0.0308823529412 & 0.0441176470588 & 0.0397058823529 & 0.0257352941176 & 0.0220588235294 \\
& {$D_\mathrm{s}(\frac{\pi}{2})$} & 0.0514705882353 & 0.0441176470588 & 0.0397058823529 & 0.0308823529412 & 0.0202205882353 \\
& {$D_\mathrm{s}(\pi)$} & 0.216176470588 & 0.158823529412 & 0.119117647059 & 0.0823529411765 & 0.0661764705882 \\
& {$R_\mathrm{P}$} & 0.0318449410867 & 0.0266484054885 & 0.0214766814517 & 0.0169099292857 & 0.0131384993739 \\
& & & & & & \\
{\num{20.0}} & {$D_\mathrm{s}(0)$} & 0.0397058823529 & 0.0470588235294 & 0.0573529411765 & 0.0647058823529 & 0.0430147058824 \\
& {$D_\mathrm{s}(\frac{\pi}{2})$} & 0.0794117647059 & 0.0735294117647 & 0.0621323529412 & 0.0566176470588 & 0.0496323529412 \\
& {$D_\mathrm{s}(\pi)$} & 0.344117647059 & 0.258823529412 & 0.191176470588 & 0.1375 & 0.1125 \\
& {$R_\mathrm{P}$} & 0.0633133927026 & 0.0535566096277 & 0.0439099473005 & 0.0364372570915 & 0.0298331532477 \\
& & & & & & \\
{\num{30.0}} & {$D_\mathrm{s}(0)$} & 0.0741176470588 & 0.0794117647059 & 0.0794117647059 & 0.0705882352941 & 0.0705882352941 \\
& {$D_\mathrm{s}(\frac{\pi}{2})$} & 0.148235294118 & 0.138970588235 & 0.125735294118 & 0.129411764706 & 0.141176470588 \\
& {$D_\mathrm{s}(\pi)$} & 0.648529411765 & 0.383823529412 & 0.357352941176 & 0.247058823529 & 0.167647058824 \\
& {$R_\mathrm{P}$} & 0.139986557805 & 0.115494655436 & 0.097696583791 & 0.0802457173131 & 0.0668162071678 \\
& & & & & & \\
{\num{40.0}} & {$D_\mathrm{s}(0)$} & 0.127205882353 & 0.112058823529 & 0.147058823529 & 0.117647058824 & 0.0941176470588 \\
& {$D_\mathrm{s}(\frac{\pi}{2})$} & 0.279852941176 & 0.242794117647 & 0.179411764706 & 0.272058823529 & 0.176470588235 \\
& {$D_\mathrm{s}(\pi)$} & 0.941323529412 & 0.616323529412 & 0.441176470588 & 0.397058823529 & 0.294117647059 \\
& {$R_\mathrm{P}$} & 0.268939566841 & 0.218783898654 & 0.182878464383 & 0.14983645943 & 0.124674031765 \\
& & & & & & \\
{\num{70.0}} & {$D_\mathrm{s}(0)$} & 0.223235294118 & 0.205882352941 & 0.22 & 0.176470588235 & 0.176470588235 \\
& {$D_\mathrm{s}(\frac{\pi}{2})$} & 1.07897058824 & 0.794117647059 & 0.6325 & 0.432352941176 & 0.367647058824 \\
& {$D_\mathrm{s}(\pi)$} & 1.11617647059 & 0.911764705882 & 0.77 & 0.564705882353 & 0.470588235294 \\
& {$R_\mathrm{P}$} & 0.701754037378 & 0.553504369983 & 0.506094681895 & 0.353094378175 & 0.303680181811 \\
& & & & & & \\
{\num{120.0}} & {$D_\mathrm{s}(0)$} & 1.17573529412 & 0.342794117647 & 0.314117647059 & 0.245588235294 & 0.224117647059 \\
& {$D_\mathrm{s}(\frac{\pi}{2})$} & 1.88117647059 & 1.17529411765 & 1.06014705882 & 0.62625 & 0.494926470588 \\
& {$D_\mathrm{s}(\pi)$} & 2.19470588235 & 1.51808823529 & 1.25647058824 & 0.810441176471 & 0.709705882353 \\
& {$R_\mathrm{P}$} & 1.68429871574 & 1.20131531766 & 1.12098718722 & 0.788414414813 & 0.642051689874 \\

\bottomrule
\end{tabular}
\end{table}

\begin{table}
\centering
\caption{Measurements of $D_\mathrm{s}(\theta)$, which is the distance from the star to the edge of the shocked stellar wind region along a direction with polar angle $\theta$. The analytical radius for a radiative bubble at the pressure confinement time, $R_\mathrm{P}$ (see \cref{equ:confine-radius} and \cref{equ:radius-rb}), is also given. These measurements were made for stars in a local number density of \SI{3.2e4}{\per\centi\meter\cubed}.}
\label{tab:sizes-mvt-shock}
\sisetup{
table-number-alignment=center,
table-text-alignment=center,
table-format=1.4e0
}
\begin{tabular}{
S[round-mode=places,round-precision=0,table-format=3.0e0]
S[table-text-alignment=left]
S[round-mode=figures,round-precision=2,table-format=1.4e0]
S[round-mode=figures,round-precision=2,table-format=1.4e0]
S[round-mode=figures,round-precision=2,table-format=1.4e0]
S[round-mode=figures,round-precision=2,table-format=1.4e0]
S[round-mode=figures,round-precision=2,table-format=1.4e0]
}
\toprule
{$M_\star$} & \multicolumn{1}{c}{Dist.} & \multicolumn{5}{c}{Age [\si{\kilo\year}]} \TopStrut{}\BottomStrut{} \\
\cline{3-7}
\multicolumn{1}{c}{[\si{\msun}]} & \multicolumn{1}{c}{[\si{\parsec}]} & {\num[round-mode=figures,round-precision=1]{20}} & {\num[round-mode=figures,round-precision=1]{40}} & {\num[round-mode=figures,round-precision=1]{60}} & {\num[round-mode=figures,round-precision=1]{80}} & {\num[round-mode=figures,round-precision=1]{100}} \TopStrut{}\BottomStrut{} \\
\midrule

{\num{6.0}} & {$D_\mathrm{s}(0)$} & 0.00882352941176 & 0.00617647058824 & 0.00588235294118 & 0.00941176470588 & 0.00941176470588 \\
& {$D_\mathrm{s}(\frac{\pi}{2})$} & 0.0111764705882 & 0.0102941176471 & 0.0102941176471 & 0.0111764705882 & 0.0111764705882 \\
& {$D_\mathrm{s}(\pi)$} & 0.0129411764706 & 0.00977941176471 & 0.0129411764706 & 0.0123529411765 & 0.0129411764706 \\
& {$R_\mathrm{P}$} & 0.00376234856381 & 0.00349792721615 & 0.00360446938063 & 0.0038663817198 & 0.00381448336842 \\
& & & & & & \\
{\num{9.0}} & {$D_\mathrm{s}(0)$} & 0.00882352941176 & 0.0105882352941 & 0.0147058823529 & 0.0105882352941 & 0.0176470588235 \\
& {$D_\mathrm{s}(\frac{\pi}{2})$} & 0.0102941176471 & 0.00860294117647 & 0.0117647058824 & 0.00823529411765 & 0.00882352941176 \\
& {$D_\mathrm{s}(\pi)$} & 0.0220588235294 & 0.0264705882353 & 0.0220588235294 & 0.00705882352941 & 0.00588235294118 \\
& {$R_\mathrm{P}$} & 0.00324035758735 & 0.00357208289145 & 0.00353360414923 & 0.00342705694604 & 0.00338803889021 \\
& & & & & & \\
{\num{12.0}} & {$D_\mathrm{s}(0)$} & 0.0176470588235 & 0.0129411764706 & 0.0282352941176 & 0.0220588235294 & 0.0441176470588 \\
& {$D_\mathrm{s}(\frac{\pi}{2})$} & 0.0161764705882 & 0.0177941176471 & 0.0211764705882 & 0.0275735294118 & 0.0275735294118 \\
& {$D_\mathrm{s}(\pi)$} & 0.05 & 0.0776470588235 & 0.0529411764706 & 0.0808823529412 & 0.0661764705882 \\
& {$R_\mathrm{P}$} & 0.00898700333613 & 0.0107122658751 & 0.0115930967723 & 0.0121161611073 & 0.0124392734343 \\
& & & & & & \\
{\num{15.0}} & {$D_\mathrm{s}(0)$} & 0.0411764705882 & 0.0376470588235 & 0.0352941176471 & 0.0423529411765 & 0.0476470588235 \\
& {$D_\mathrm{s}(\frac{\pi}{2})$} & 0.0329411764706 & 0.0376470588235 & 0.0411764705882 & 0.0388235294118 & 0.0436764705882 \\
& {$D_\mathrm{s}(\pi)$} & 0.0494117647059 & 0.0988235294118 & 0.141176470588 & 0.176470588235 & 0.190588235294 \\
& {$R_\mathrm{P}$} & 0.016968883182 & 0.0210551001903 & 0.0233940495363 & 0.025422348295 & 0.0269952431828 \\
& & & & & & \\
{\num{20.0}} & {$D_\mathrm{s}(0)$} & 0.0352941176471 & 0.0588235294118 & 0.0573529411765 & 0.0435294117647 & 0.0470588235294 \\
& {$D_\mathrm{s}(\frac{\pi}{2})$} & 0.0411764705882 & 0.0588235294118 & 0.08125 & 0.0652941176471 & 0.0588235294118 \\
& {$D_\mathrm{s}(\pi)$} & 0.105882352941 & 0.183823529412 & 0.229411764706 & 0.282941176471 & 0.329411764706 \\
& {$R_\mathrm{P}$} & 0.030956208816 & 0.0407418933286 & 0.0464683552554 & 0.0519116790848 & 0.0550375099295 \\
& & & & & & \\
{\num{30.0}} & {$D_\mathrm{s}(0)$} & 0.0917647058824 & 0.0705882352941 & 0.0794117647059 & 0.0882352941176 & 0.0735294117647 \\
& {$D_\mathrm{s}(\frac{\pi}{2})$} & 0.0879411764706 & 0.1 & 0.119117647059 & 0.117647058824 & 0.117647058824 \\
& {$D_\mathrm{s}(\pi)$} & 0.183529411765 & 0.329411764706 & 0.370588235294 & 0.411764705882 & 0.514705882353 \\
& {$R_\mathrm{P}$} & 0.0666028503246 & 0.0881434632273 & 0.10189406469 & 0.114753511277 & 0.123546363979 \\
& & & & & & \\
{\num{40.0}} & {$D_\mathrm{s}(0)$} & 0.123529411765 & 0.147058823529 & 0.147058823529 & 0.114705882353 & 0.141176470588 \\
& {$D_\mathrm{s}(\frac{\pi}{2})$} & 0.205882352941 & 0.176470588235 & 0.205882352941 & 0.210294117647 & 0.258823529412 \\
& {$D_\mathrm{s}(\pi)$} & 0.277941176471 & 0.411764705882 & 0.411764705882 & 0.65 & 0.729411764706 \\
& {$R_\mathrm{P}$} & 0.122067531394 & 0.116797778741 & 0.191948228801 & 0.214246030677 & 0.233690804276 \\
& & & & & & \\
{\num{70.0}} & {$D_\mathrm{s}(0)$} & 0.176470588235 & 0.223529411765 & 0.194117647059 & 0.202941176471 & 0.229411764706 \\
& {$D_\mathrm{s}(\frac{\pi}{2})$} & 0.323529411765 & 0.558823529412 & 0.695588235294 & 0.642647058824 & 0.707352941176 \\
& {$D_\mathrm{s}(\pi)$} & 0.470588235294 & 0.698529411765 & 0.808823529412 & 0.879411764706 & 1.07058823529 \\
& {$R_\mathrm{P}$} & 0.286323245764 & 0.456252941391 & 0.525114562617 & 0.61467342105 & 0.682421965588 \\
& & & & & & \\
{\num{120.0}} & {$D_\mathrm{s}(0)$} & 0.323529411765 & 0.294117647059 & 0.264705882353 & 0.329411764706 & 0.370588235294 \\
& {$D_\mathrm{s}(\frac{\pi}{2})$} & 0.355882352941 & 0.900735294118 & 1.05882352941 & 1.22352941176 & 1.37647058824 \\
& {$D_\mathrm{s}(\pi)$} & 0.582352941176 & 1.13970588235 & 1.32352941176 & 1.41176470588 & 1.58823529412 \\
& {$R_\mathrm{P}$} & 0.774576933508 & 0.983113640094 & 1.11087394466 & 1.21703764276 & 1.3478609779 \\

\bottomrule
\end{tabular}
\end{table}

\Cref{tab:sizes-mvd-shock,tab:sizes-mvt-shock} list the sizes of the shocked stellar wind regions as a function of density and time respectively.
In \cref{tab:sizes-mvt-shock} the sizes of the shocked stellar wind regions show a plateau at early times for low-mass stars.
The plateau occurs at later times for higher-mass stars.
This suggests that the bubbles are becoming pressure-confined (plots of the pressure confirm this).

We use the results of an analysis of stellar wind evolution by \citet{1992ApJ...388...93K} to see if this behaviour is physical.
The terminal wind velocity in all of the models is lower than the critical wind velocity defined in \citet[see their equation 2.5]{1992ApJ...388...93K}, indicating that the wind-bubbles are radiative.
In this regime the cooling time of the shocked wind is shorter than the time it takes to accumulate a significant mass of gas ahead of the shock.
We found that the radii of the shocked stellar wind regions are consistent with pressure-confined fully radiative bubbles.

Numerical diffusion can cause extra cooling, i.e. the cooling length might not be sufficiently resolved and therefore intermediate temperatures are found in a higher volume of the gas.
Kelvin-Helmholtz instabilities can also enhance cooling by increasing the surface area of the contact discontinuity.
Increasing the resolution of the numerical grid increases the growth of these instabilities to further enhance cooling.
These behaviours are competing but overall the level of numerical diffusion has been found to decrease with higher resolution \citep{2010MNRAS.406.2373P}.
We carried out resolution tests that confirm that the cooling across the contact discontinuity was not enhanced significantly by numerical effects.

The shell radius for a radiative bubble is given by
\begin{equation}
\label{equ:radius-rb}
R_\mathrm{shell} = \left(\frac{3 \mathscr{L}}{\pi \rho_\mathrm{i} v_\infty}\right)^{1/4} t^{1/2} \mathpunc{,}
\end{equation}
where $\mathscr{L} = \frac{1}{2} \dot{M} v_\infty^2$ is the mechanical wind luminosity and $\rho_\mathrm{i}$ is the ionised ambient density.
For stars with $M_\star \geq \SI{70}{\msun}$ the wind is driving a shock that follows \cref{equ:radius-rb} just ahead of the ionisation front, which explains why we are seeing large ionisation front radii in \cref{tab:sizes-mvd-if,tab:sizes-mvt-if} for these stars.
The shock ahead of the ionisation front for the lower-mass stars evolves according to \cref{equ:raga-1}, indicating that the wind-blown bubble has been pressure-confined by the \hii{} region.

A radiative bubble eventually becomes pressure-confined when the ram pressure of the shell is equal to the ambient pressure.
Using \cref{equ:radius-rb} we can therefore obtain the time at which confinement occurs:
\begin{equation}
\label{equ:confine-time}
t_\mathrm{P} = \left(\frac{3 \mathscr{L}}{16 \pi \rho_\mathrm{i} c_\mathrm{i}^4 v_\infty}\right)^{1/2} \mathpunc{,}
\end{equation}
where $c_\mathrm{i}$ is the isothermal sound speed in the ionised ambient medium.
The ionised ambient density, $\rho_\mathrm{i}$, was determined for each snapshot as the density upstream of the shock in the ionised ambient region in the radial direction.
Using \cref{equ:ion-balance} we can get an approximation for this density:
\begin{equation}
\label{equ:confine-radius}
\rho_\mathrm{i} = \left(\frac{R_\mathrm{st}}{D_\mathrm{i}(\tfrac{\pi}{2})}\right)^{3/2} \rho_\star \mathpunc{,}
\end{equation}
where $D_\mathrm{i}(\frac{\pi}{2})$ is the measured radial distance from star to ionisation front given in \cref{tab:sizes-mvd-if,tab:sizes-mvt-if} ($D_\mathrm{i}(\frac{\pi}{2})$ gives an ``average'' value for the bubble radius).

For most of the models the transition time is too short to be seen either because the earliest data snapshot was taken at \SI{2}{\kilo\year} or because the bubble is small and therefore not well resolved.
As a result the bubbles quickly stall and consequently are missing the shock outside of the contact discontinuity that separates shocked stellar wind material and the ionised ambient medium.
As the shock is missing and the wind-blown bubbles are behaving like radiative bubbles, the radial distance from the star to the contact discontinuity, $D_\mathrm{s}(\frac{\pi}{2})$ in \cref{tab:sizes-mvd-shock,tab:sizes-mvt-shock}, should be comparable to the analytical radius of the shock at the confinement time given in \cref{equ:confine-time}.
The approximate final radius of the shocked wind region in the radial direction
is
\begin{equation}
R_\mathrm{P} = R_\mathrm{shell}(t_\mathrm{P}) = \left(\frac{3 \mathscr{L}}{4 \pi \rho_\mathrm{i} c_\mathrm{i}^2 v_\infty}\right)^{1/2} \mathpunc{,}
\end{equation}
which is also given for each model in \cref{tab:sizes-mvd-shock,tab:sizes-mvt-shock}.
Looking at these tables, the measured transverse distances, $D_\mathrm{s}(\frac{\pi}{2})$, show closer agreement with the confinement radius, $R_\mathrm{P}$, for higher-mass stars.
In \cref{tab:sizes-mvd-shock,tab:sizes-mvt-shock}, we can see generally better agreement for stars in lower density environments and at later times respectively.
These results suggest that stars with high enough mass have stellar winds that dominate the expansion of the \hii{} region.
For lower-mass stars the shocked stellar wind region is noticeably larger than predicted, suggesting that the expansion of the \hii{} region is also contributing to the expansion of the shocked stellar wind region.

In all of the models the contact discontinuity eventually stalls.
For stars with mass $\SI{6}{\msun} \le M_\star \le \SI{30}{\msun}$ in \cref{tab:sizes-mvt-shock} the stellar wind size in the transverse direction, $D_\mathrm{s}(\frac{\pi}{2})$, plateaus before they reach an age of \SI{100}{\kilo\year}.
This behaviour is easily seen for the star with mass $M_\star = \SI{30}{\msun}$ but is harder to see for lower-mass stars because the contact discontinuity between the shocked stellar wind region and the ionised ambient region is unstable.

\subsection{Emission Measures}

\begin{figure*}
	\centering
	\includegraphics[width=0.7\linewidth]{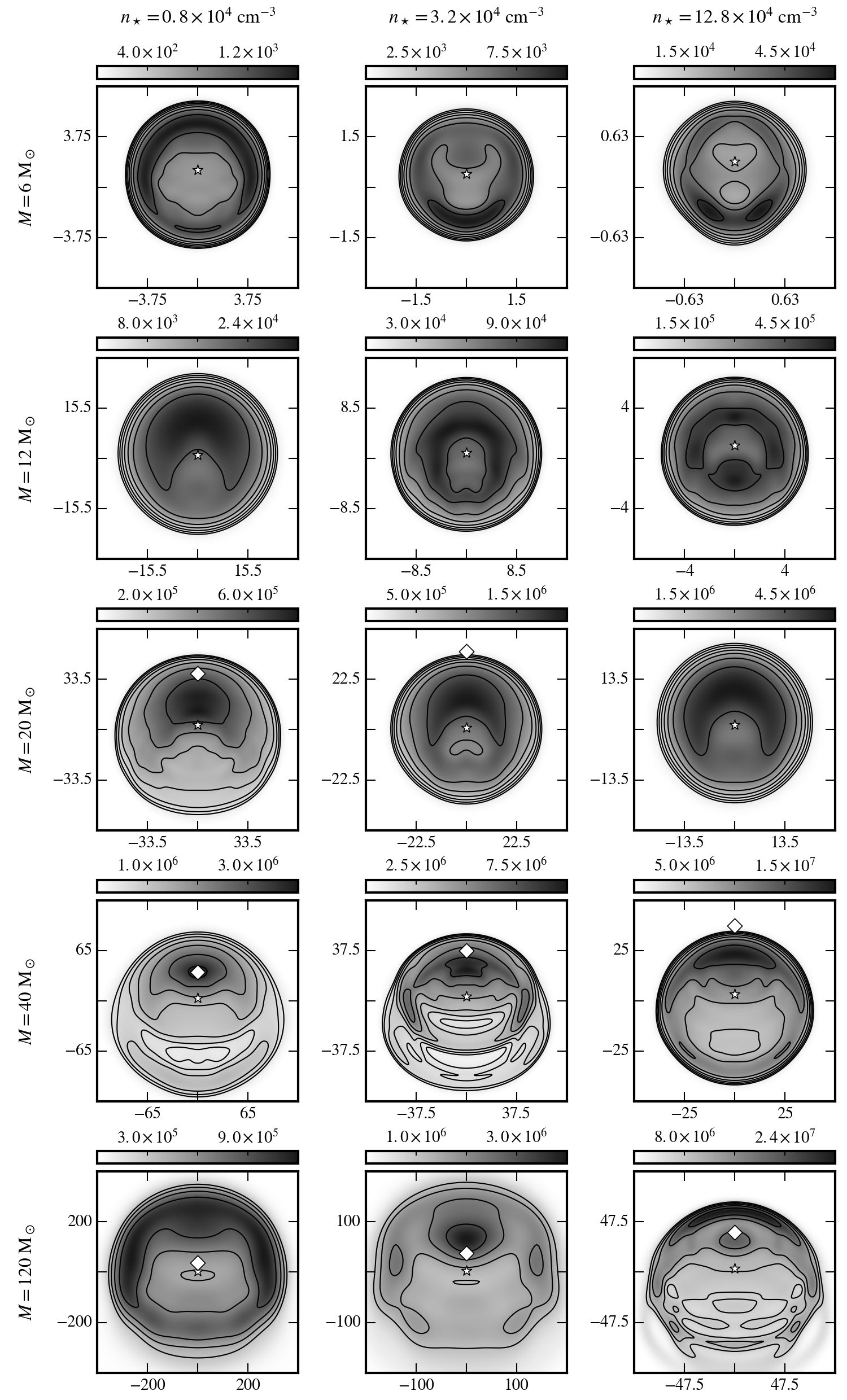}
	\caption{The emission measures (in \si{\parsec\per\centi\meter\tothe{6}}) of simulated \gls*{uchii} regions at an age of $t = \SI{50000}{\year}$ and viewed at a projection angle (see \cref{equ:projection-angle}) of $\theta_\mathrm{i} = \SI{45}{\degree}$. Each row shows the emission measure for a star of a specified stellar mass (given on the far left) at increasing local number densities going right (number densities are given at the top of each column). The axes are in units of arcseconds and the object is assumed to be at a distance of \SI{1.5}{\kilo\parsec} from the observer. Each map also shows logarithmic contours at $\sqrt{2}$ intervals from the maximum emission measure. The star marker on each plot shows the position of the star and the diamond marker (where visible) shows the position of the cloud centre.}
	\label{fig:em-mass-density}
\end{figure*}
\begin{figure*}
	\centering
	\includegraphics[width=0.7\linewidth]{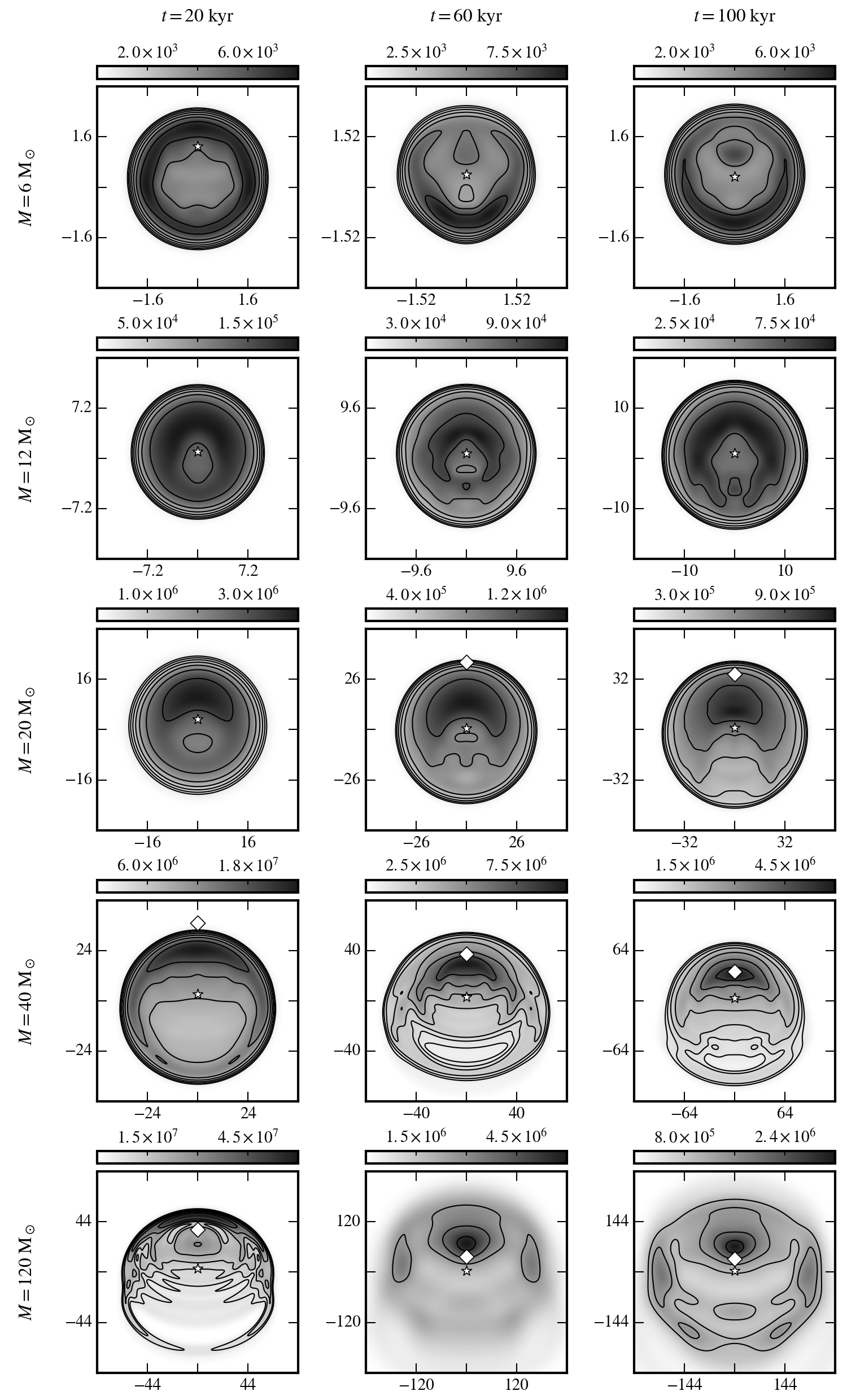}
	\caption{The emission measures (in \si{\parsec\per\cm\tothe{6}}) of simulated \gls*{uchii} regions in a local hydrogen number density of $n_\star = \SI{3.2e4}{\per\centi\meter\cubed}$ and viewed at a projection angle (see \cref{equ:projection-angle}) of $\theta_\mathrm{i} = \SI{45}{\degree}$. Each row shows the emission measure for a star of a specified stellar mass (given on the far left) at increasing ages going right (age is given at the top of each column). The axes are in units of arcseconds and the object is assumed to be at a distance of \SI{1.5}{\kilo\parsec} from the observer. Each map also shows logarithmic contours at $\sqrt{2}$ intervals from the maximum emission measure. The star marker on each plot shows the position of the star and the diamond marker (where visible) shows the position of the cloud centre.}
	\label{fig:em-mass-time}
\end{figure*}

In order to see what our models would look like if the ionised region is optically thin we calculated the emission measure:
\begin{equation}
EM = \int n_\mathrm{e}^2 \mathpunc{} ds \mathpunc{,}
\end{equation}
where $n_\mathrm{e}$ is the number density of free electrons and the integration is along a line of sight.

To do this integration we used the cylindrically symmetric ray-tracing scheme presented in the appendix of \citet{2003A&A...409..217D}.
Grids of the resulting emission measures at a viewing projection angle of \SI{45}{\degree} are shown spanning two dimensional parameter spaces in \cref{fig:em-mass-density,fig:em-mass-time}. 
In \cref{fig:em-mass-density} the effects of varying stellar mass and the star's local density were explored at a time of \SI{50}{\kilo\year}. 
The evolution over time of emission measures for different stellar masses is plotted in \cref{fig:em-mass-time} at a local number density of $n_\star = \SI{3.2e4}{\per\centi\metre\cubed}$.

The figures clearly show two distinct behaviours.
For stars with mass $M_\star \leq \SI{9}{\msun}$ (and also $M_\star = \SI{12}{\msun}$ for $n_\star > \SI{3.2e4}{\per\centi\metre\cubed}$) the morphology of the cavity blown out by the stellar wind varies over the simulation time.
In these models the stellar wind is weak and is therefore isotropically confined by the \hii{} region, explaining the non-cometary morphologies.
A pressure gradient exists in these \hii{} regions, which leads to the flow of material down the density gradient.
Stars with higher mass have wind bubbles that are not confined down the density gradient, leading to expansion in this direction.
These stars have cometary morphologies that do not change appreciably over a period of \SI{200}{\kilo\year} and show limb-brightening.
Stars with masses $M_\star \geq \SI{70}{\msun}$ also show limb-brightening if $n_\star > \SI{3.2e4}{\per\centi\metre\cubed}$.
The rest have \hii{} regions that expand past the centre of the dense cloud so that this is the only feature picked up in the emission measure images.
For the \hii{} region in the bottom left of \cref{fig:em-mass-density} we see limb-brightening because the high density core is destroyed by the stellar wind.

\begin{figure*}
	\includegraphics[width=1.00\linewidth]{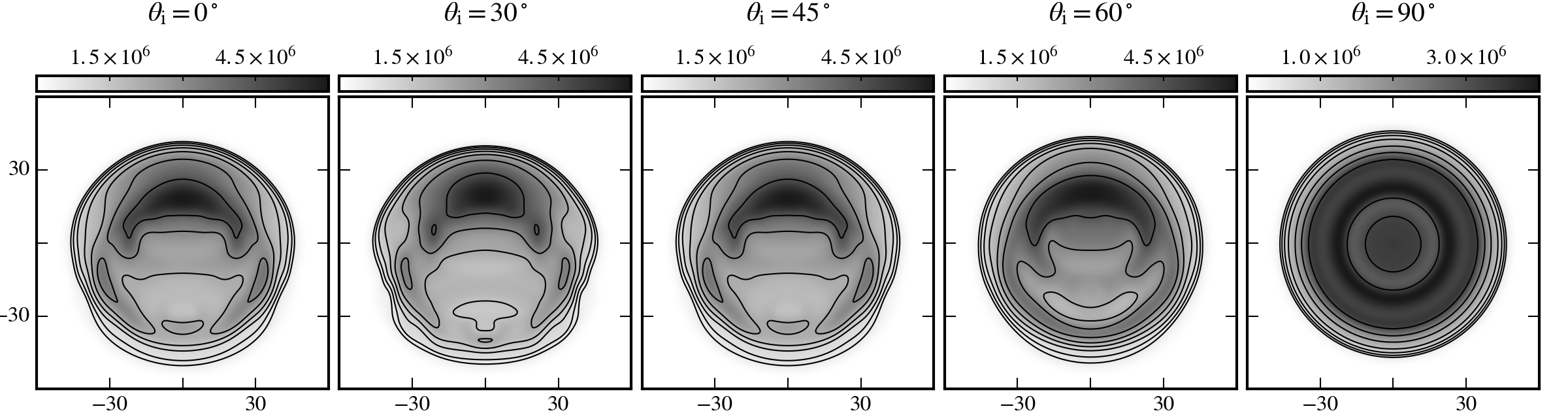}
	\caption{The emission measures (in \si{\parsec\per\cm\tothe{6}}) of a simulated \hii{} region around a star with mass $M_\star = \SI{30}{\msun}$ in a local hydrogen number density of $n_\star = \SI{3.2e4}{\per\centi\meter\cubed}$ at an age of $t = \SI{50}{\kilo\year}$ viewed at different projection angles (see \cref{equ:projection-angle}), which are given at the top of each plot. The axes are in units of arcseconds and the object is assumed to be at a distance of \SI{1.5}{\kilo\parsec} from the observer.}
	\label{fig:emeasure-angle}
\end{figure*}

For one of the models that does exhibit limb-brightening we show in \cref{fig:emeasure-angle} the effects of changing the projection angle on the morphology.
The projection angle is defined as
\begin{equation}
\label{equ:projection-angle}
\theta_\mathrm{i} = \arccos \left( \hat{z} \customdot \hat{d} \right) \mathpunc{,}
\end{equation}
where $\hat{z}$ is a unit vector directed from star to cloud centre and $\hat{d}$ is a unit vector directed from star to observer.
At projection angles closer to $\SI{90}{\degree}$ limb-brightening is more pronounced i.e. the ``limbs'' wrap further around the centre of the object.
Using the classification scheme of \citet{2005ApJ...624L.101D} the projected morphology is less cometary and more shell-like the closer the axis of symmetry of the \hii{} region is oriented towards the observer.
These results suggest that \hii{} regions classified as having a shell-like morphology may be cometaries viewed along their axis of symmetry \cite[as noted by][]{1991ApJ...369..395M}.

\subsection{Spectral Indices}

\begin{figure}
	\includegraphics[width=1.00\linewidth]{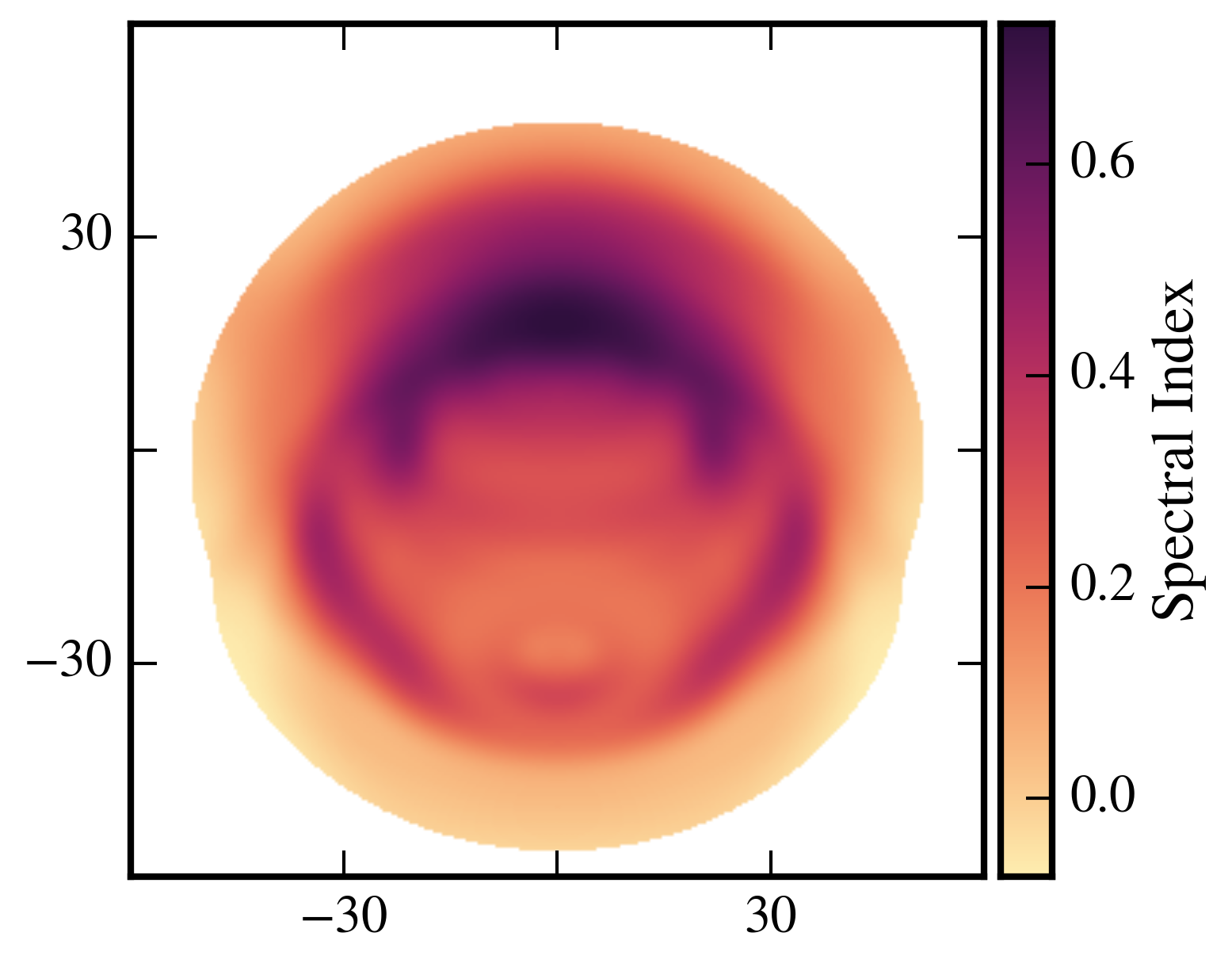}
	\caption{A spectral index map of radio continuum emission from a model \hii{} region around a star of mass $M_\star = \SI{30}{\msun}$ in a local hydrogen number density of $n_\star = \SI{3.2e4}{\per\centi\meter\cubed}$ and at an age of $t = \SI{50}{\kilo\year}$. The object was viewed at a projection angle (see \cref{equ:projection-angle}) of $\theta_\mathrm{i} = \SI{45}{\degree}$. Spectral indices for each pixel were calculated using two flux density images of the model for frequencies of \SI{1.4}{\giga\hertz} and \SI{5.0}{\giga\hertz}. The axes are in units of arcseconds and the object is assumed to be at a distance of \SI{1.5}{\kilo\parsec} from the observer. The region in white is below the noise level of the \gls*{cornish} survey (\SI{\sim 0.4}{\milli\janskey\per\beam}) therefore a spectral index was not calculated. The integrated spectral index of this image is $\alpha = 0.36$ (see \cref{tab:spectral-indices-mvd}).}
	\label{fig:spectral-index-map}
\end{figure}

\begin{table}
\centering
\caption{Spectral indices of model \hii{} regions at an age of \SI{50}{\kilo\year} and viewing projection angle (see \cref{equ:projection-angle}) of $\theta_\mathrm{i} = \SI{45}{\degree}$. The spectral indices were calculated using total fluxes for each model at frequencies of \SI{1.4}{\giga\hertz} and \SI{5.0}{\giga\hertz}.}
\label{tab:spectral-indices-mvd}
\sisetup{
table-number-alignment = center,
table-format=5.4e0,
table-text-alignment = center
}
\begin{tabular}{
S[round-mode=places,round-precision=0,table-format=3.0e0]
S[round-mode = places,round-precision = 2,table-format=-1.2e0]
S[round-mode = places,round-precision = 2,table-format=-1.2e0]
S[round-mode = places,round-precision = 2,table-format=-1.2e0]
S[round-mode = places,round-precision = 2,table-format=-1.2e0]
S[round-mode = places,round-precision = 2,table-format=-1.2e0]
}
\toprule
{$M_\star$} & \multicolumn{5}{c}{$\mathrm{n_\star}$ [\SI[retain-unity-mantissa=false]{1e4}{\per\centi\meter\cubed}]} \TopStrut{}\BottomStrut{} \\
\cline{2-6}
{[\si{\msun}]} & {\num[round-mode=places,round-precision=1]{0.8}} & {\num[round-mode=places,round-precision=1]{1.6}} & {\num[round-mode=places,round-precision=1]{3.2}} & {\num[round-mode=places,round-precision=1]{6.4}} & {\num[round-mode=places,round-precision=1]{12.8}} \TopStrut{}\BottomStrut{} \\
\midrule

6.0 & -0.106015732359 & -0.105482006428 & -0.104391881487 & -0.102706179757 & -0.0980196814551 \\
& & & & & \\
9.0 & -0.104571395621 & -0.103617434232 & -0.100887836101 & -0.093570599772 & -0.0734433549647 \\
& & & & & \\
12.0 & -0.100842264895 & -0.096975044268 & -0.0896366307692 & -0.0738250807688 & -0.0331017178063 \\
& & & & & \\
15.0 & -0.0856683428401 & -0.0732099996997 & -0.051443460897 & -0.00688607982049 & 0.0739193046948 \\
& & & & & \\
20.0 & -0.0264370609143 & 0.0201737342144 & 0.10175959701 & 0.214497133977 & 0.383359438423 \\
& & & & & \\
30.0 & 0.0772901962985 & 0.195171090881 & 0.364911177448 & 0.560428164982 & 0.833907761373 \\
& & & & & \\
40.0 & 0.145898142881 & 0.30388479943 & 0.493877162408 & 0.763856735207 & 1.03729603886 \\
& & & & & \\
70.0 & 0.122398064313 & 0.297548019315 & 0.36275381179 & 0.784283533384 & 1.07782880331 \\
& & & & & \\
120.0 & 0.0358526022948 & 0.158910455649 & 0.274016887296 & 0.661190627486 & 1.05320217143 \\

\bottomrule
\end{tabular}
\end{table}

\begin{table}
\centering
\caption{Spectral indices of model \hii{} regions in a local hydrogen number density of $n_\star = \SI{3.2e4}{\per\centi\meter\cubed}$ and viewing projection angle (see \cref{equ:projection-angle}) of $\theta_\mathrm{i} = \SI{45}{\degree}$.}
\label{tab:spectral-indices-mvt}
\sisetup{
table-number-alignment = center,
table-format=5.4e0,
table-text-alignment = center
}
\begin{tabular}{
S[round-mode=places,round-precision=0,table-format=3.0e0]
S[round-mode = places,round-precision = 2,table-format=-1.2e0]
S[round-mode = places,round-precision = 2,table-format=-1.2e0]
S[round-mode = places,round-precision = 2,table-format=-1.2e0]
S[round-mode = places,round-precision = 2,table-format=-1.2e0]
S[round-mode = places,round-precision = 2,table-format=-1.2e0]
}
\toprule
{$M_\star$} & \multicolumn{5}{c}{Age [\si{\kilo\year}]} \TopStrut{}\BottomStrut{} \\
\cline{2-6}
{[\si{\msun}]} & {\num[round-mode=figures,round-precision=1]{20}} & {\num[round-mode=figures,round-precision=1]{40}} & {\num[round-mode=figures,round-precision=1]{60}} & {\num[round-mode=figures,round-precision=1]{80}} & {\num[round-mode=figures,round-precision=1]{100}} \TopStrut{}\BottomStrut{} \\
\midrule

6.0 & -0.10473914627 & -0.104451755312 & -0.104482427616 & -0.104761612702 & -0.104789397562 \\
& & & & & \\
9.0 & -0.0996088756973 & -0.100905638888 & -0.100780594367 & -0.100523221696 & -0.100278311698 \\
& & & & & \\
12.0 & -0.0765832331897 & -0.0877786624268 & -0.0911182104782 & -0.0927639418507 & -0.0933029545403 \\
& & & & & \\
15.0 & 0.00962177358359 & -0.0434208495814 & -0.0586936144425 & -0.0672420963741 & -0.0717381331554 \\
& & & & & \\
20.0 & 0.345933585819 & 0.136817018362 & 0.0663211680502 & 0.0299936138482 & 0.00302654922364 \\
& & & & & \\
30.0 & 0.824879752297 & 0.453102037733 & 0.291525959879 & 0.189392629785 & 0.136750678526 \\
& & & & & \\
40.0 & 1.10162980075 & 0.615286650834 & 0.426011321152 & 0.309101729436 & 0.228168069663 \\
& & & & & \\
70.0 & 1.13245486841 & 0.461388921878 & 0.315408260463 & 0.221488819135 & 0.167599915983 \\
& & & & & \\
120.0 & 0.93776193326 & 0.382275685955 & 0.214960943009 & 0.140240896765 & 0.095797749511 \\

\bottomrule
\end{tabular}
\end{table}

The change in specific intensity, at frequency $\nu$, across a ray segment in a grid cell is
\begin{equation}
	\Delta I_\nu = \left( 1 - e^{-\Delta\tau_\nu} \right) \left(\frac{j_\nu}{\alpha_\nu} - I_0 \right) \mathpunc{,}
\end{equation}
where $I_0$ is the specific intensity at the start of the path segment, $j_\nu$ and $\alpha_\nu$ are the emission and absorption coefficients respectively, and $\Delta\tau_\nu$ is the optical depth across the segment given by
\begin{equation}
\Delta\tau_\nu = \alpha_\nu \Delta s \mathpunc{,}
\end{equation}
where $\Delta s$ is the path length through the cell.
The emission coefficient \citep{1979rpa..book.....R} is
\begin{equation}
j_\nu = \frac{32 \pi q^6 k_\mathrm{e}^3}{3m_\mathrm{e}^{3/2}c^3} \sqrt{\frac{2\pi}{3k_\mathrm{B} T}} Z^2 n_\mathrm{e} n_\mathrm{i} e^{- h\nu / k_\mathrm{B} T} g_{\mathrm{ff}} \mathpunc{,}
\end{equation}
where $q$ is the elementary charge, $k_\mathrm{e}$ is Coulomb's constant, $m_\mathrm{e}$ is the mass of an electron, $Z$ is the ionic charge, $h$ is Planck's constant, and $g_{\mathrm{ff}}$ is the velocity averaged Gaunt factor \citep{1988ApJ...327..477H}.
Using Kirchoff's law,
\begin{equation}
j_\nu = \alpha_\nu B_\nu (T) \mathpunc{,}
\end{equation}
where $B_\nu (T)$ is the Planck function, we have the absorption coefficient:
\begin{equation}
\alpha_{\nu} = \frac{16 \pi q^6 k_\mathrm{e}^3}{3m_\mathrm{e}^{3/2}ch\nu^3} \sqrt{\frac{2\pi}{3k_\mathrm{B} T}} Z^2 n_\mathrm{e} n_\mathrm{i} \left(1 - e^{- h\nu / k_\mathrm{B} T}\right) g_{\mathrm{ff}} \mathpunc{.}
\end{equation}

We produced synthetic radio continuum maps by accumulating the changes in intensity along rays through the finite cells in our hydrodynamic solutions to pixels on an image plane.
The flux density, $S_\nu$, in each pixel was obtained by integrating the intensity over the solid angle of the pixel.
Spectral indices were approximated using these maps:
\begin{equation}
\mathrm{Spectral\ Index} = \frac{\partial \log S_\nu}{\partial \log \nu} \simeq \frac{\log S_{\nu_0} - \log S_{\nu_1}}{\log \nu_0 - \log \nu_1} \mathpunc{,}
\end{equation}
where $\nu_0$ and $\nu_1$ are two closely separated frequencies.

A spectral index map for radio continuum emission is plotted in \cref{fig:spectral-index-map} for a star in the centre of the parameter space we are exploring for frequencies \SI{1.4}{\giga\hertz} and \SI{5.0}{\giga\hertz}.
Intermediate spectral indices (between \num{-0.1} and \num{+2}) are found across the whole image.
At the head and the limbs of the \hii{} region, where the emission measure is highest, the spectral indices are higher (\num{\sim 0.6}) than those in the cavity blown out by the stellar wind (\num{\sim 0.2}).

\Cref{tab:spectral-indices-mvd} shows the spectral indices of models for a range of local densities and masses.
These indices were calculated using the total radio continuum flux of each model for frequencies of \SI{1.4}{\giga\hertz} and \SI{5.0}{\giga\hertz}.
At these frequencies the objects are more optically thick the higher the stellar mass (this is also seen in \cref{tab:spectral-indices-mvt}).
Higher-mass stars have more intense ionising radiation and blow stronger winds, so they ionise deeper into the dense cloud, leading to a higher proportion of the total flux from areas of high density.
For masses $M_\star > \SI{40}{\msun}$ the spectral indices decrease, which is likely due to the competing effect of larger regions having a higher proportion of flux from optically thin regions.
These stars are ionising the regions around the densest part of the cloud (or through it in the case of the star with $M_\star = \SI{120}{\msun}$ in a local density of $n_\star = \SI{12.8e4}{\per\centi\meter\cubed}$). 

In \cref{tab:spectral-indices-mvt} spectral indices are shown to decrease over time.
The density in a \hii{} region decreases as it expands (after its initial Str\"{o}mgren expansion phase), so the spectral indices should decrease also.

\begin{figure}
	\includegraphics[width=1.00\linewidth]{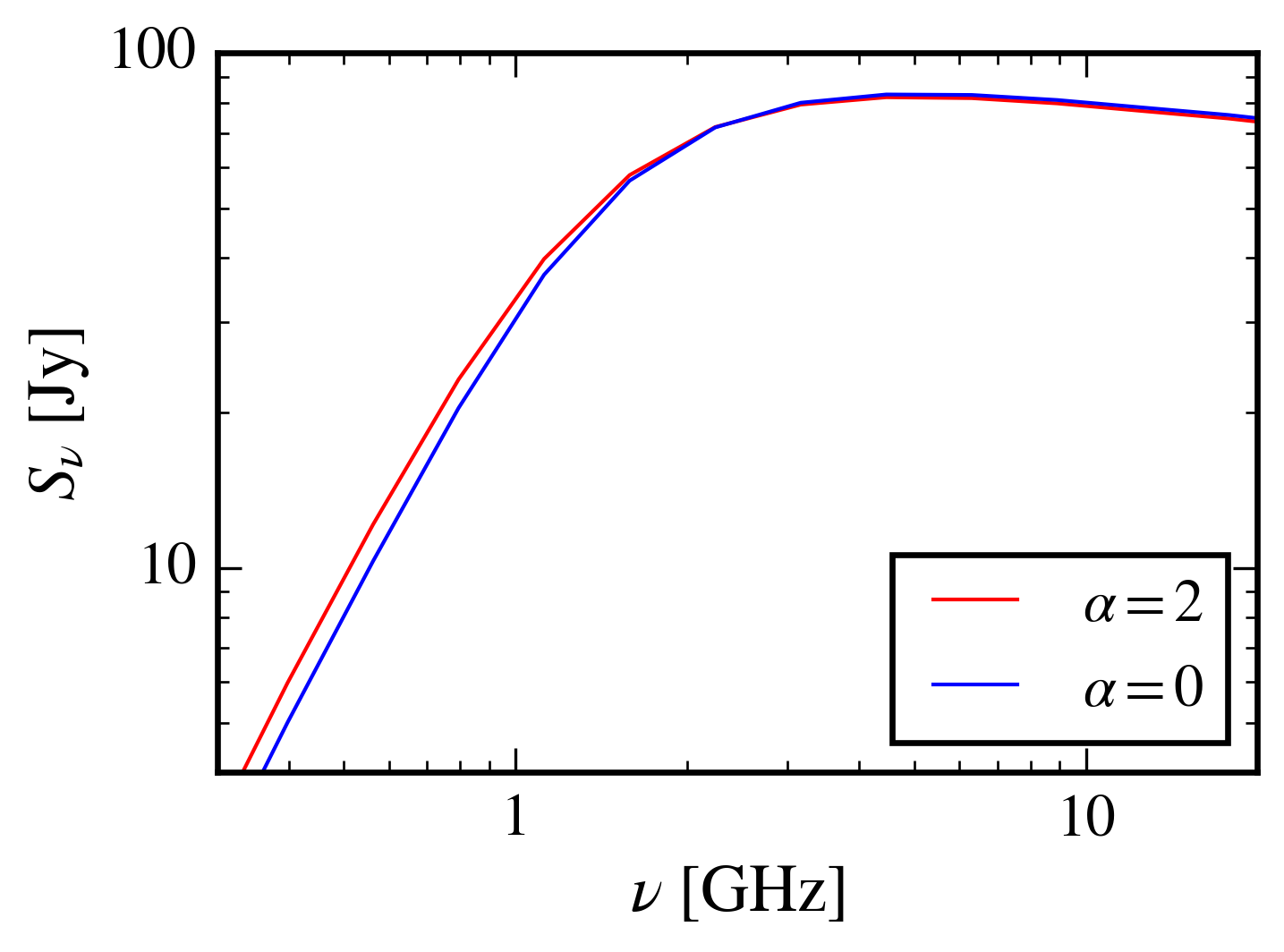}
	\caption{The radio spectrum of two model \hii{} regions around stars of mass $M_\star = \SI{30}{\msun}$ with a local hydrogen number density of $n_\star = \SI{3.2e4}{\per\centi\meter\cubed}$ at an age of $t = \SI{50}{\kilo\year}$ and at a projection angle (see \cref{equ:projection-angle}) of $\theta_\mathrm{i} = \SI{45}{\degree}$. One of the stars is in a power-law density environment, $\alpha = 2$, and the other is in a uniform density environment, $\alpha = 0$. Both objects were viewed at a projection angle of $\theta_\mathrm{i} = \SI{45}{\degree}$ and assumed to be at a distance of \SI{1.5}{\kilo\parsec}.}
	\label{fig:sed}
\end{figure}

\Cref{fig:sed} shows the radio spectrum of one of our model \hii{} regions and also a \hii{} region that is produced by the same star but in a uniform density environment.
Surprisingly, the turn-over frequency is very similar for the power-law case, but it is slightly lower because the \hii{} region covers more low-density gas down the slope than high-density gas up the slope.
The total fluxes have a higher proportion from lower density gas and therefore the integrated spectral indices will lean closer to indices from lower density regions.
The close similarity between the SEDs is consistent with the fact that the density slope of ionised gas is much shallower than the initial slope in the neutral ambient gas.

\subsection{Model Limitations}

If we had included the effects of the entire spectrum of radiation from our model stars the temperature structures of our \hii{} regions would be different.
We would see higher temperatures near the ionisation front due to the hardening of the radiation field \citep{2004MNRAS.353.1126W}.
The extra pressure due to these high temperatures is expected to be \SI{\sim 20}{\percent} higher than in our model \hii{} regions.

Another limitation of our model is that we took the initial thermal pressure of the neutral gas to be uniform to prevent gas from moving away from the cloud centre (which would happen if the temperature was uniform instead). 
In a future study the effects of gravity could be included to better model these environments.

Dust is an important factor in the evolution of high density \hii{} regions that we did not consider.
Ionising photons can be absorbed by dust and therefore dusty \hii{} regions are expected to be smaller.
\citet{2004ApJ...608..282A} simulated dusty compact \hii{} regions and found that the fraction of ionising photons absorbed by dust decreases as the \hii{} region expands, and so the effects of dust are less important in the later phases of expansion.
They also found that dusty \hii{} regions stagnate earlier than their dust-free counterparts.

\section{Conclusions}
\label{sec:conclusions}

We have simulated cometary \hii{} regions that result from stars that blow stellar winds and are placed off-centre from a power-law density gas cloud.
Models with stars of mass $M_\star \geq \SI{12}{\msun}$ produce limb-brightening.
Some are diminished if the star's \hii{} region contains the dense centre of the cloud (which is true for stars with high Lyman output near low density clouds).
There is a cavity blown out by the stellar wind in each model, which is observed in real cometary \hii{} regions.
The morphological class of a \hii{} region depends on the viewing projection angle.
Morphologies are more shell-like the smaller the angle between the viewing direction and the object's axis of symmetry.
Spectral indices are higher for higher-mass stars, which can be explained as larger \hii{} regions covering more of the high density gas near the cloud centre.
This effect turns over at even higher masses as the volume of ionised low density gas increases more than the volume of ionised high density gas in the cloud centre.
Model spectral energy distributions for power-law density models were almost identical to those of the same stars placed in uniform density surroundings, which is consistent with the shallow density structures we found in our simulated ionised regions.
The turn-over frequency in the radio spectrum for our power-law density models is slightly lower compared to the models with uniform ambient density.
This is because \hii{} regions in a power-law density medium cover more low density gas than high density gas, which means the proportion of emission from optically thin regions is higher.

In a subsequent paper we will use the grid of simulated models presented in this paper to select \hii{} regions that correspond to regions generated in a model Milky Way.
The ultimate goal is to compare observables from this simulated survey with the \gls*{cornish} survey in order to investigate how well our models describe \hii{} region evolution.

\section*{Acknowledgements}

We thank the anonymous reviewer for their critical reading of this paper and for their suggestions.
This work was supported by the Science and Technology Facilities Council at the University of Leeds.





\bibliographystyle{mnras}
\DeclareRobustCommand{\VAN}[3]{#3}
\bibliography{references} 
\DeclareRobustCommand{\VAN}[3]{#2}



\appendix

\section{Code Tests}
\label{app:code-tests}

\subsection{R-Type Expansion and Shadowing}

Two tests were carried out to ensure the accuracy of the radiative transfer scheme. 
The first test compares the radius of a simulated Str\"omgren sphere with the analytical solution for the Str\"omgren radius \citep{1939ApJ....89..526S}:
\begin{equation}
	R_\mathrm{st} = 
	\left(\frac{3 Q_\mathrm{Lyc}}{4 \pi n_{\mathrm H}^2 \alpha_{\mathrm B}}\right)^{1/3}\left(1 - e^{-t/t_{\mathrm{rec}}}\right)^{1/3} \mathpunc{,}
	\label{equ:stromgren-tdep}
\end{equation}
where $t_\mathrm{rec} = 1 / n_\mathrm{H} \alpha_\mathrm{B}$ is the recombination time.
We assume here that $Q_\mathrm{Lyc} = \SI[retain-unity-mantissa=false]{1e48}{\photons\per\second}$, $n_\mathrm{H} = \SI{100}{\per\centi\meter\cubed}$, and $\alpha_\mathrm{B} = \SI{2.59e-13}{\centi\meter\cubed\per\second}$.

\begin{figure}
	\centering
	\includegraphics[width=\linewidth]{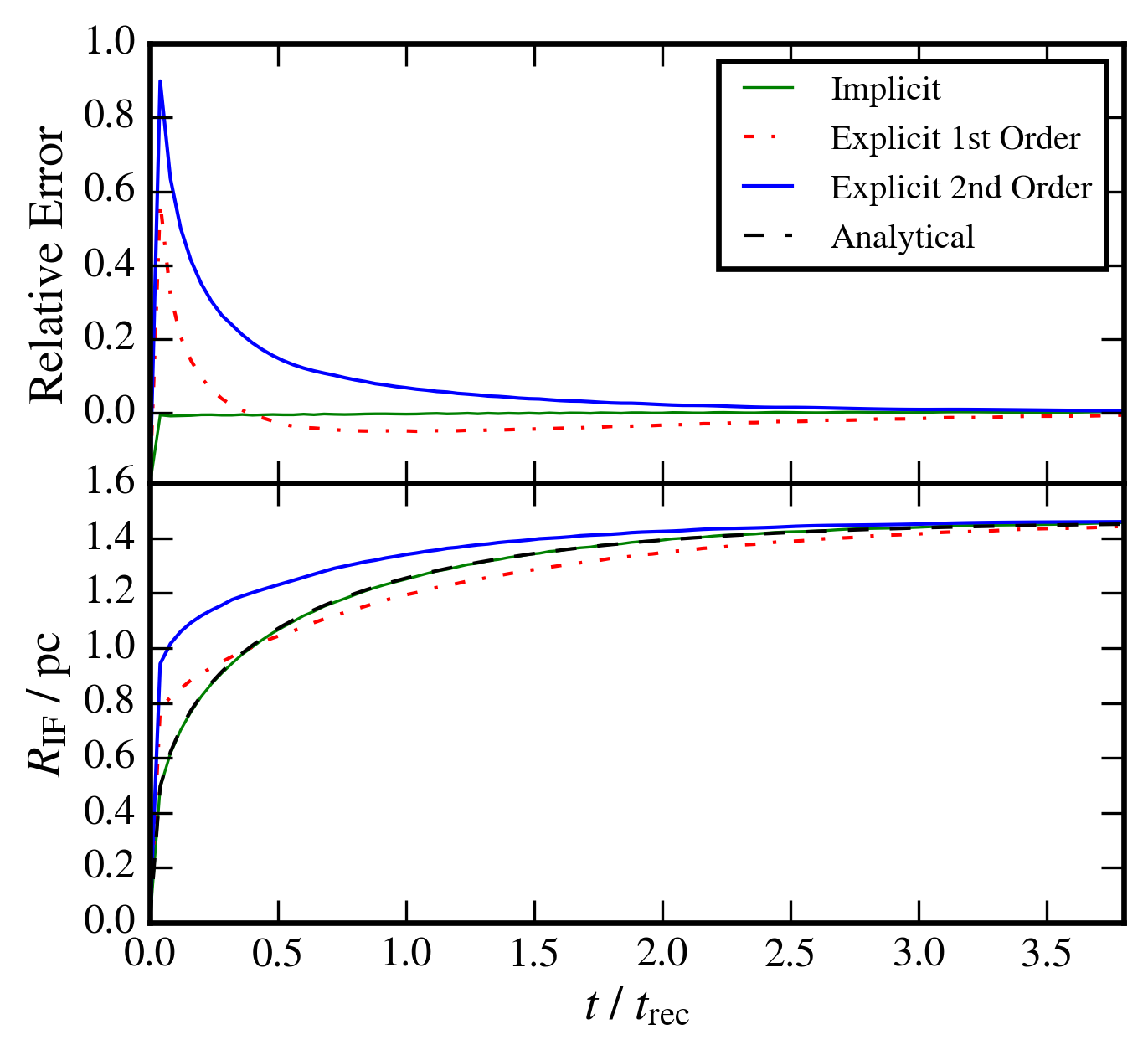}
	\caption{Expansion of a \hii{} region as a function of time in a uniform density environment. There is no coupling to the hydrodynamics. The dashed black curve represents the analytical solution, which is partially obscured by the solid green curve (simulated with the implicit scheme). The ionisation fronts simulated by explicit first-order and second-order schemes are shown by the dash-dotted red curve and the thick solid blue curve respectively.}
	\label{fig:stromgren-tdep}
\end{figure}

The results of this test are shown in \cref{fig:stromgren-tdep} along with a comparison between first-order and second-order explicit schemes and the implicit scheme \gls*{torch} uses.
Clearly the implicit scheme out-performs the other schemes, converging to the analytical solution in a time much smaller than the recombination time.

\begin{figure}
	\centering
	\includegraphics[width=1.0\linewidth]{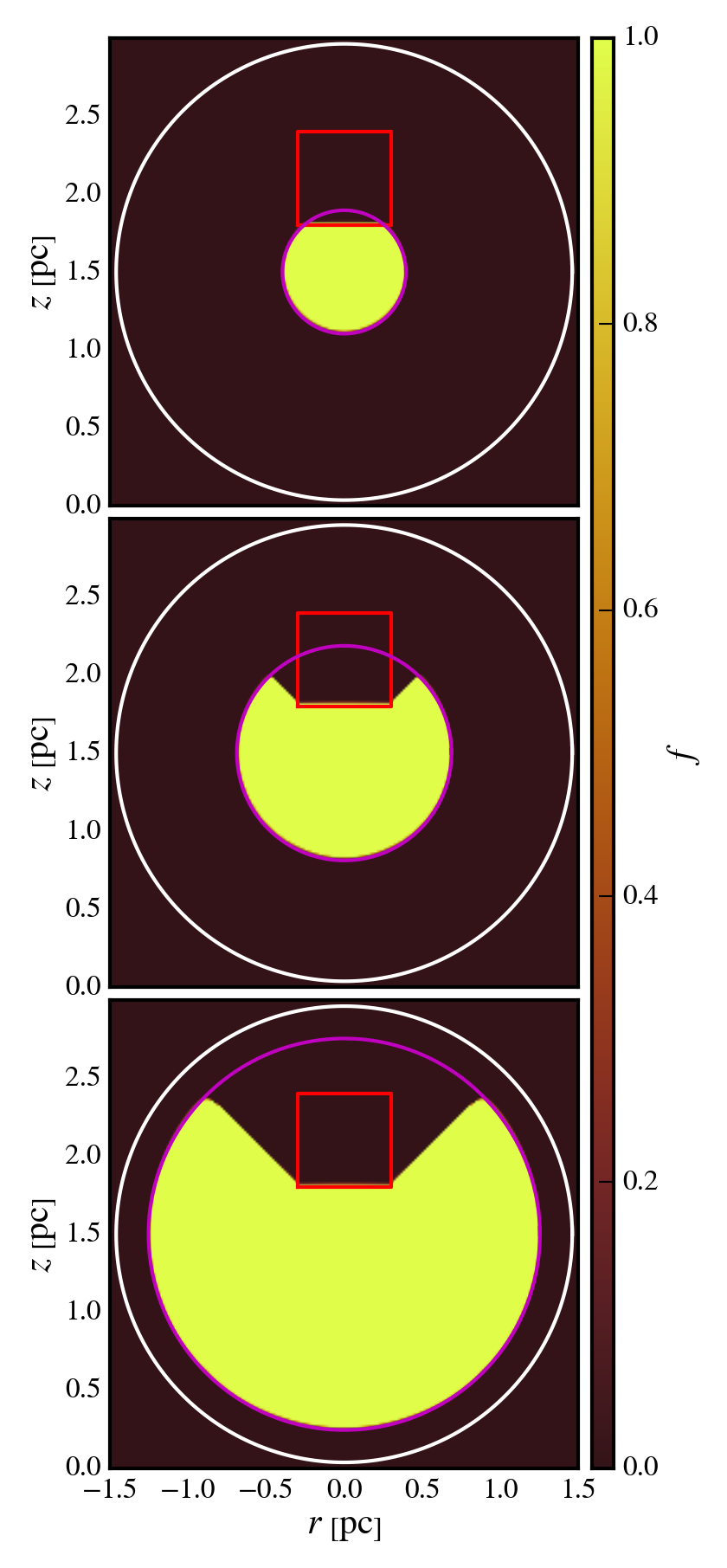}
	\caption{\hii{} fraction evolution in a uniform density medium with a dense square clump (in red) nearby. The magenta circle traces the analytical Str\"omgren sphere, which approaches the white circle as $t \rightarrow \infty$. The snapshots were taken at \SI{24}{\year} (top), \SI{135}{\year} (middle) and \SI{1224}{\year} (bottom). The star was positioned at $(\SI{0}{\parsec}, \SI{1.5}{\parsec})$ and the centre of the square clump (of side-length \SI{0.6}{\parsec}) was positioned at $(\SI{0}{\parsec}, \SI{2.1}{\parsec})$.}
	\label{fig:shadow}
\end{figure}

Simulations of \gls*{uchii} regions with stellar winds should show trapping of the ionisation front behind the swept-up wind bubble.
To see if the code works accurately in such a situation, the Str\"omgren sphere test was repeated but with a dense square clump positioned near the ionising source.
The clump effectively blocks photo-ionising radiation from propagating into the ``shadow'' region.
There is, however, some numerical diffusion into the shadow region due to the interpolative nature of the short-characteristics ray-tracing scheme.
This will have a negligible effect on the dynamics of any simulated \hii{} region.
\Cref{fig:shadow} shows the evolution of the ionised \hii{} fraction in the shadow test.

\subsection{D-Type Expansion}
\label{sec:dtype}

\begin{figure}
	\centering
	\includegraphics[width=1.0\linewidth]{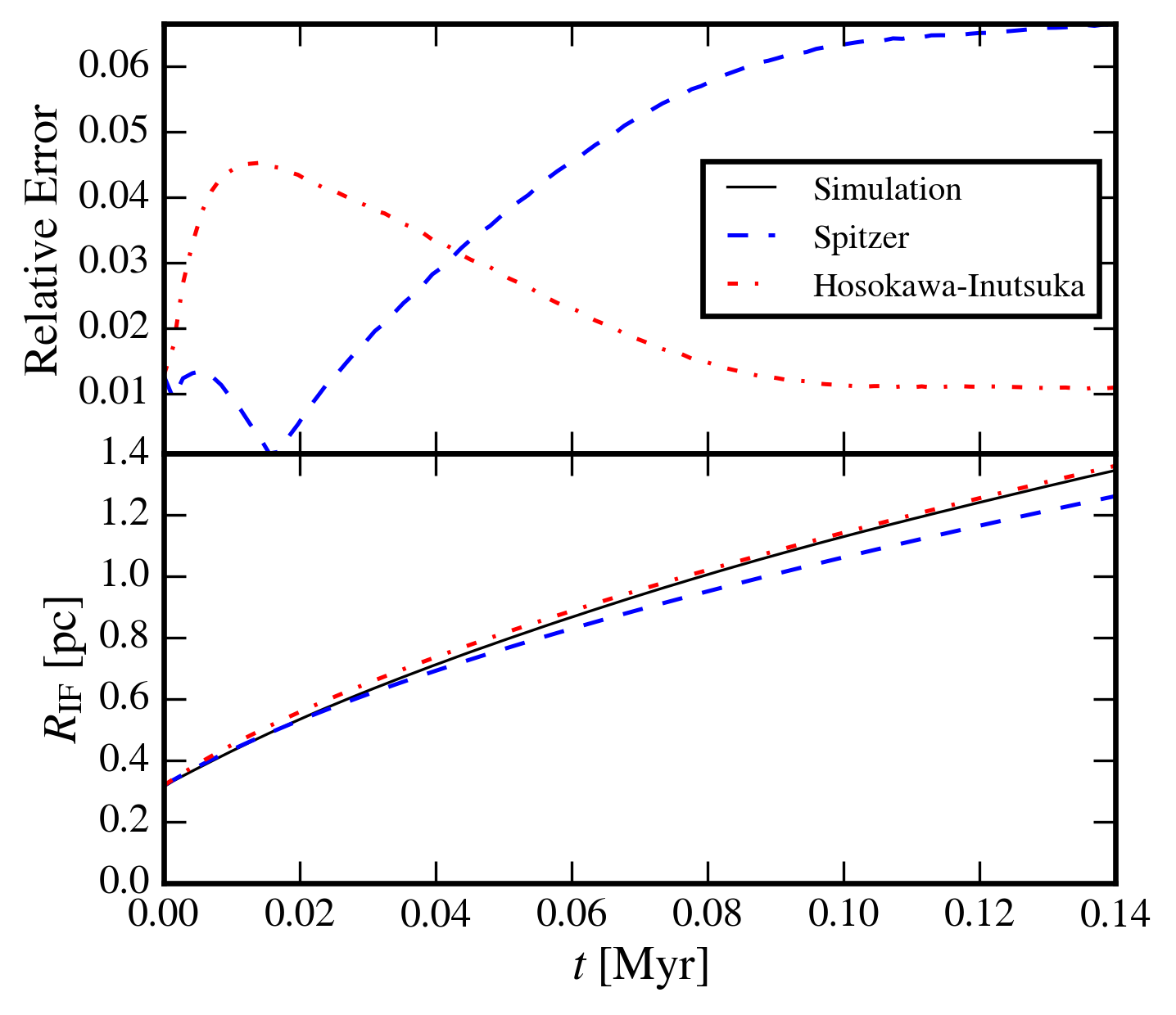}
	\caption{The one-dimensional simulation results (using \num{2000} cells) of the early-phase expansion of a \hii{} region in a uniform density environment described in \cref{sec:dtype}. The solid black curve shows the simulated ionisation front radius. The Spitzer (\cref{equ:spitzer}) and Hosokawa-Inutsuka (\cref{equ:hosokawa-if}) solutions are given by the dashed blue curve and dash-dotted red curve respectively.}
	\label{fig:early}
\end{figure}

\begin{figure}
	\centering
	\includegraphics[width=1.0\linewidth]{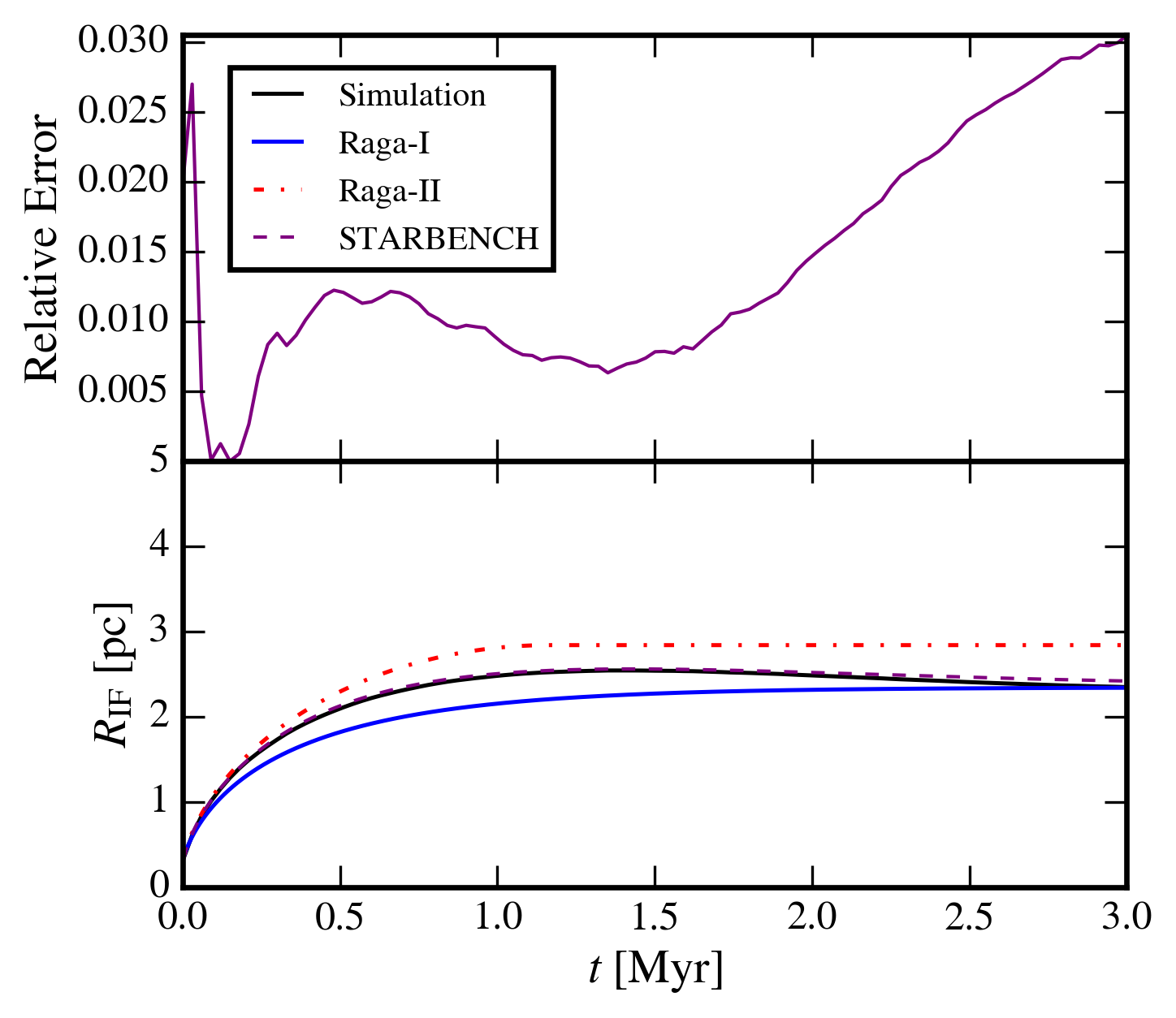}
	\caption{The one-dimensional simulation results (using \num{2000} cells) of the late-phase expansion of a \hii{} region in a uniform density environment described in \cref{sec:dtype}. The solid black curve shows the simulated ionisation front radius. The analytical radii labelled Raga-I (thick solid blue curve) and Raga-II (dash-dotted red curve) show the analytical radii predicted by \cref{equ:raga-1} and \cref{equ:raga-2} respectively.}
	\label{fig:late}
\end{figure}

\gls*{torch} was benchmarked against the results of the one-dimensional spherically symmetric D-type expansion tests of \citet{2015MNRAS.453.1324B}.
The early and late phase simulations, described by this author, of D-type ionisation front expansion into an initially uniform density ($\rho_\mathrm{H} = \SI{5.21e-21}{\gram\per\centi\meter\cubed}$) isothermal gas were run.
The ionising star in this simulation has a Lyman continuum photon rate of $Q_\mathrm{Lyc} = \SI[retain-unity-mantissa=false]{1e49}{\photons\per\second}$, which produces an initial Str\"{o}mgren sphere of radius $R_\mathrm{st} = \SI{0.314}{\parsec}$ and temperature $T_\mathrm{HII} = \SI{10000}{\kelvin}$.
The temperature of the neutral ambient medium was set to $T_\mathrm{HI} = \SI{100}{\kelvin}$ for the early phase and $T_\mathrm{HI} = \SI{1000}{\kelvin}$ for the late phase.
Coupling between gas and radiation physics was achieved for an isothermal equation of state by setting the temperature of a gas element according to the ionised fraction of hydrogen in that element:
\begin{equation}
	T = T_\mathrm{HI} + (T_\mathrm{HII} - T_\mathrm{HI}) f_\mathrm{HII} \mathpunc{,}
\end{equation}
where $T_\mathrm{HI}$ is the temperature of fully neutral hydrogen and $T_\mathrm{HII}$ is the temperature of fully ionised hydrogen.

Early on in the evolution (a few recombination times) a weak-R ionisation front expands according to \cref{equ:stromgren-tdep}. 
The gas is heated so that the ionised region is over-pressured with respect to the neutral gas.
As the ionisation front slows down it becomes R-critical; the pressure wave overtakes the front and steepens into a shock wave, compressing the gas behind it.
The R-critical ionisation front transitions into a D-critical front with an isothermal shock ahead of it.
D-type expansion is then driven by the over-pressure on the sound crossing time-scale $t_\mathrm{s} = R_\mathrm{st} / c_\mathrm{i}$, where $R_\mathrm{st}$ is the Str\"{o}mgren radius given in \cref{equ:stromgren-tdep} and $c_\mathrm{i} = \SI{12.85}{\kilo\meter\per\second}$ is the sound speed of the ionised gas.
\Cref{fig:early} shows the expansion of the ionisation front in the early phase.
The expansion begins by following the Spitzer solution (\cref{equ:spitzer}) and eventually follows the solution by \citet{2006ApJ...646..240H} (\cref{equ:hosokawa-if}).
In \cref{fig:late}, we show the ionisation front evolution in the late phase.
The simulated front is a close fit to the S\textsc{tar}B\textsc{ench} equation \citep{2015MNRAS.453.1324B}, with an error of less than \SI{3}{\percent} for the simulation duration.

\subsection{Wind-Blown Bubble}
\label{sec:wind-bubble-app}

The wind-blown bubble model of \citet{1998MNRAS.297..747S} was simulated using \gls*{torch} to test if the cooling instability in the shell is reproduced.
A stellar wind with mass loss rate $\dot{M}_W = \SI{5e-5}{\msun\per\year}$ and terminal wind speed $v_W = $ \SI{2000}{\kilo\metre\per\second} was modelled by including source terms $\dot{\rho}_W$ and $\dot{e}_W$ within a radius of 10 grid cells (see \cref{sec:stellar-winds}).

\begin{figure}
	\centering
	\includegraphics[width=1.0\linewidth]{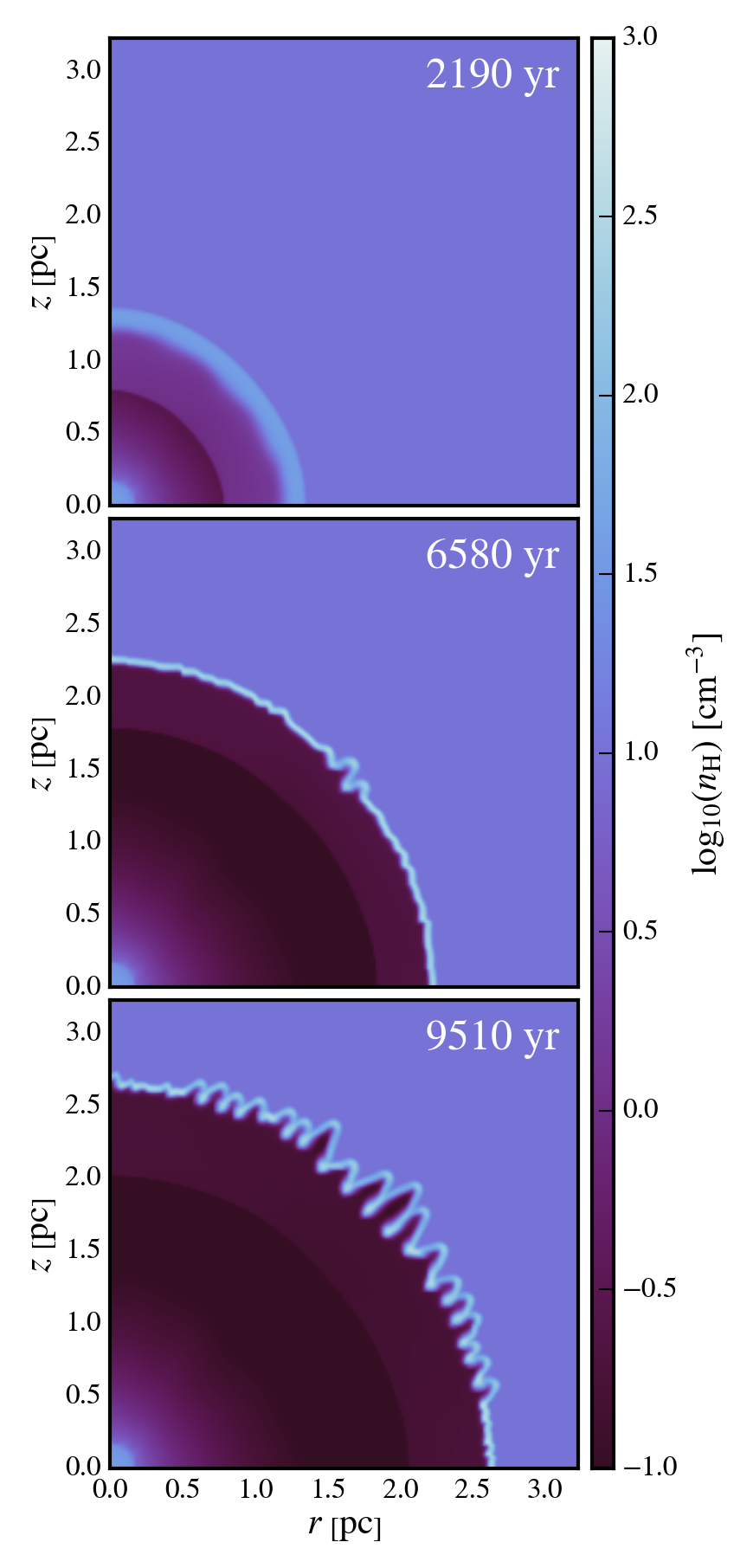}
	\caption{Snapshots of $\mathrm{log}_{10}$ of the hydrogen number density $n_\mathrm{H}$ taken during a simulation of a wind-blown bubble.}
	\label{fig:strickland}
\end{figure}

As was found in \citet{1998MNRAS.297..747S} the wind sweeps material into a shock-heated shell.
The shell starts out thick, but rapidly cools and collapses as it grows.
Eventually the shell is so thin that it becomes unstable to thin shell and Vishniac instabilities \citep{1983ApJ...274..152V}.
The evolution of the ionised hydrogen number density during the simulation is shown in \cref{fig:strickland} and agrees well with the model in \citet{1998MNRAS.297..747S}.

\subsection{Shadowing Instability}
This final test utilises both the radiation and cooling modules in order to reproduce the shadowing instability \citep{1999MNRAS.310..789W}.
A wind was set up in the same way as in \cref{sec:stellar-winds} with wind, star and initial gas parameters taken from the shadowing instability test in \citet{2006ApJS..165..283A}.
If the cooling time is too short the dense swept-up shell can collapse while the effects of the staggered grid are still appreciable (i.e. the shell has a low resolution).
For this reason the energy flux from the heating and cooling module was artificially reduced by a factor of \num{100} in order to increase the cooling length and so delay the onset of the cooling instability.

\begin{figure}
	\centering
	\includegraphics[width=1.0\linewidth]{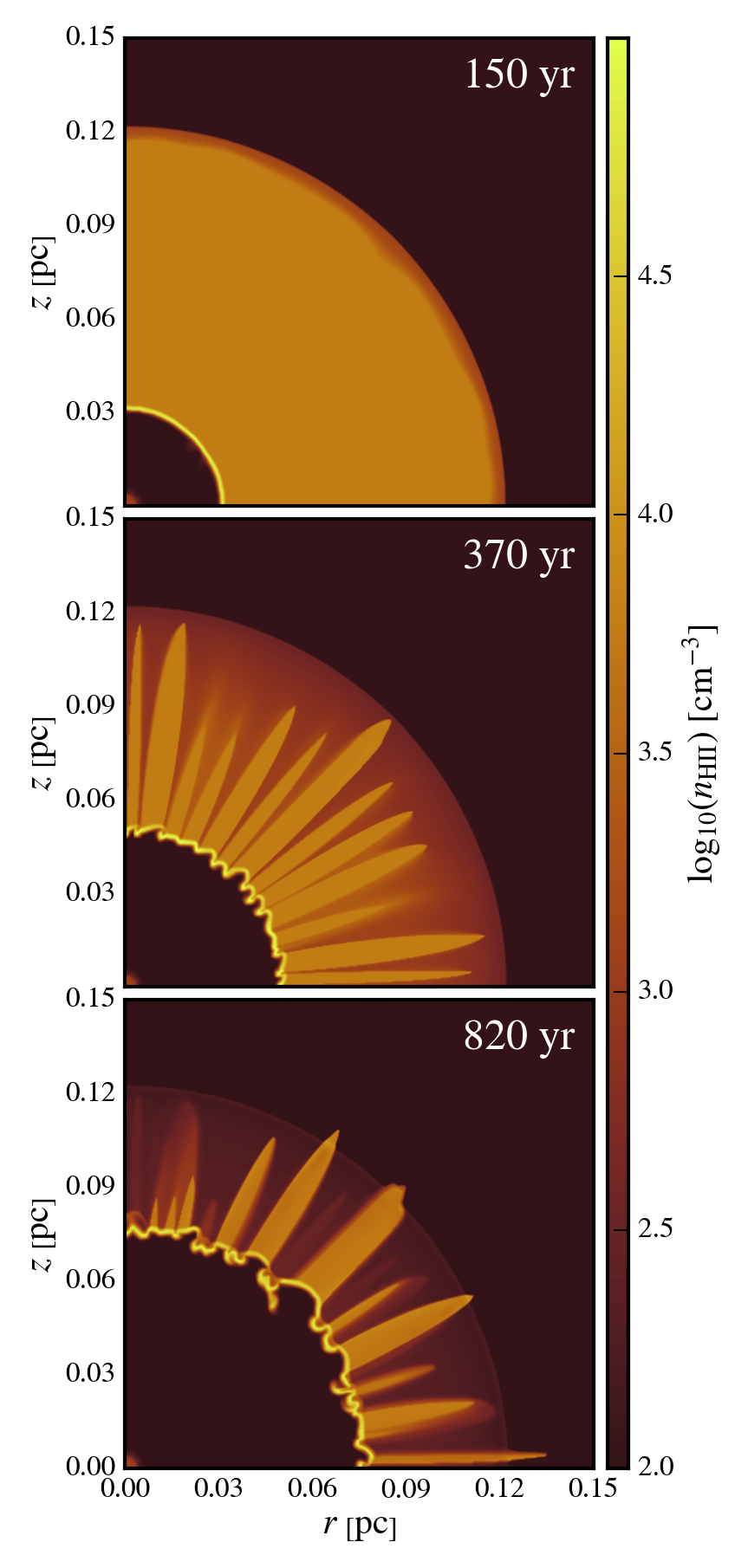}
	\caption{Snapshots of $\mathrm{log}_{10}$ of the ionised hydrogen number density $n_\mathrm{H}$ taken during a simulation of a photo-ionised wind-blown bubble.}
	\label{fig:spokes}
\end{figure}

\Cref{fig:spokes} shows the evolution of the ionised gas density \citep[similar to figure 2 in][]{2006ApJS..165..283A}, which behaves as expected.
The cooling instability discussed in \cref{sec:wind-bubble-app} occurs and produces a variation in the optical depth, leading to the tell-tale spokes of the shadowing instability.

\section{Rotated Hybrid HLL/HLLC}
\label{sec:hll-rot}

To solve the Riemann problem across inter-cell boundaries we first implemented a \gls*{hllc} Riemann solver.
After testing it was apparent that this scheme is not robust as unacceptable flow fields can result called carbuncle instabilities \citep{FLD:FLD1650180603}.
Schemes such as \gls*{hll} and Rusanov do not have this problem as they are highly dissipative.
There are a few ways to cure schemes that are prone to the carbuncle phenomenon; the technique we decided to use was to combine \gls*{hll} and \gls*{hllc} schemes into a rotated-hybrid Riemann solver \citep{Nishikawa20082560}.

The scheme starts by decomposing the geometric grid cell face normal $\customvec{n}$ into two orthogonal directions:
\begin{equation}
	\customvec{n}_1 =
	\label{eq:n1}
	\begin{cases}
		\quad \dfrac{\Delta \customvec{q}}{|\Delta \customvec{q}|}, & \quad \text{if } |\Delta \customvec{q}| > \epsilon \mathpunc{,} \\
		\quad \customvec{n}_\perp, & \quad \text{otherwise},
	\end{cases}
\end{equation}
and
\begin{equation}
	\customvec{n}_2 = \dfrac{(\customvec{n}_1 \times \customvec{n}) \times \customvec{n}_1}{|(\customvec{n}_1 \times \customvec{n}) \times \customvec{n}_1|} \mathpunc{,}
\end{equation}
where $\Delta \customvec{q} = (u_\mathrm{R}-u_\mathrm{L},v_\mathrm{R}-v_\mathrm{L})$ is the velocity difference vector, $\customvec{n}_\perp$ is a direction tangent to the geometric face and $\epsilon$ is a small number.
The second case in \cref{eq:n1} ensures that only the \gls*{hllc} solver is used when streamwise velocity fields are smoothly varying; only at discontinuities will the \gls*{hll} solver be applied. 
The flux across this grid cell face is
\begin{equation}
	\customvec{F} = \customvec{n} \customdot \customvec{n}_1 \customvec{F}_\mathrm{HLL}(\customvec{n}_1) + \customvec{n} \customdot \customvec{n}_2 \customvec{F}_\mathrm{HLLC}(\customvec{n}_2) \mathpunc{.}
	\label{eq:rot_riemann}
\end{equation}

Fluxes $\customvec{F}_\mathrm{HLL}(\customvec{n}_1)$ and $\customvec{F}_\mathrm{HLLC}(\customvec{n}_2)$ are calculated by first finding the velocity in a new  coordinate system:
\begin{equation}
	\pvec{v}' = \custommat{R} \customvec{v} \mathpunc{,}
\end{equation}
where $\customvec{v}$ is the velocity in the original coordinate system on either the left or right side of the cell interface and $\custommat{R}$ is the rotation matrix (defined as rotating decomposed Riemann problem directions, $\customvec{n}_1$ or $\customvec{n}_2$, to align with the geometric cell face normal, $\customvec{n}$).

Using the Rodrigues' rotation formula \citep{rodriguesrotation1840} the rotation matrix $\custommat{R}$, which rotates unit vector $\customvec{n}_i$ onto unit vector $\customvec{n}$ is given by:
\begin{equation}
	\custommat{R} = \custommat{I} + [\customvec{w}]_{\times} + \frac{1 - c}{s^2} [\customvec{w}]_{\times}^2 \mathpunc{,}
\end{equation}
where
\begin{gather}
	\customvec{w} = \customvec{n}_i \times \customvec{n} \mathpunc{,} \\
	c = \customvec{n}_i \customdot \customvec{n} \mathpunc{,} \\
	s = ||\customvec{w}|| \mathpunc{,}
\end{gather}
and $[\customvec{w}]_{\times}$ is the skew-symmetric cross-product matrix of $\customvec{w}$,
\begin{equation}
	[\customvec{w}]_{\times} = \begin{pmatrix}
							0 & -w_3 & w_2 \\
							w_3 & 0 & -w_1 \\
							-w_2 & w_1 & 0
						\end{pmatrix} \mathpunc{.}
\end{equation}

With rotated left and right Riemann states for each decomposed Riemann direction we can now solve the one dimensional Riemann problem. Solving these problems leaves us with momentum fluxes $\pvec{L}'$ (calculated using $\pvec{v}'$) that need to be rotated back onto the original coordinate system:
\begin{equation}
	\customvec{L} = \custommat{R}^{-1} \pvec{L}' \mathpunc{.}
\end{equation}
Along with the rest of the flux components, this gives us $\customvec{F}_\mathrm{HLL}(\customvec{n}_1)$ and $\customvec{F}_\mathrm{HLLC}(\customvec{n}_2)$.
Using the rotated Riemann solver effectively applies the \gls*{hll} solver in the direction normal to shocks (suppressing the carbuncle instability) and applies the \gls*{hllc} solver across shear layers in order to minimise dissipation.

\citet{liou2000mass} argues that intermediate cells across a shock exchange information transversely to their neighbours along the shock, which can develop into a carbuncle instability.
Considering a face connecting two neighbouring intermediate cells, any significant perturbation travelling in a direction normal to this face will lead to a velocity difference vector oriented in the same direction \citep{Nishikawa20082560}.
This means the \gls*{hll} Riemann solver will introduce dissipation in this direction, suppressing the instability.
The resulting scheme pays for its robustness with an acceptable drop in accuracy (using a dissipative Riemann solver) and speed (using two Riemann solvers).


\bsp	

\label{lastpage}
\end{document}